\documentclass[sigconf, nonacm]{acmart}

\renewcommand\footnotetextcopyrightpermission[1]{}

\usepackage{lipsum}
\usepackage{setspace}
\usepackage{verbatim}
\usepackage{comment}
\usepackage{blindtext}

\usepackage{amsfonts}       
\usepackage{amsthm}         
\usepackage{amsmath}        
\usepackage{mathtools}      
\usepackage{nicefrac}       
\usepackage{stmaryrd}       
\usepackage{mathrsfs}       
\usepackage{etoolbox}       

\usepackage{array}
\usepackage{booktabs}       
\usepackage{multirow}       

\usepackage{microtype}

\usepackage{enumerate}
\usepackage{enumitem} 

\usepackage{graphicx}   
\usepackage{caption}    
\usepackage{subcaption} 
\usepackage{float}      

\usepackage{url}

\usepackage{algorithm}
\usepackage{algpseudocodex} 

\usepackage{siunitx}

\usepackage{tikz}
\usetikzlibrary{decorations.text,calc,arrows.meta,backgrounds,shapes,arrows,patterns}
\usepackage{pgfplots}
\pgfplotsset{compat=1.18}

\usepackage{xcolor}
\usepackage{color}  
\definecolor{darkgreen}{rgb}{0.0, 0.5, 0.0}

\usepackage{pifont}   
\usepackage{csquotes} 
\usepackage{eurosym}  
\usepackage{balance}  
\usepackage{qtree}    

\theoremstyle{plain}
\newtheorem*{theorem*}{Theorem}

\usepackage{chngcntr}

\counterwithin*{weakness}{section}

\counterwithin*{detail}{section}

\counterwithin*{revision}{section}

\theoremstyle{definition}

\counterwithin*{weakness}{section}

\newcommand{\indep}{\perp \!\!\! \perp}
\newcommand{\dep}{\perp \!\!\! \not\perp}
\newcommand{\schele}[1]{\texttt{#1}}

\newcommand\vldbdoi{10.14778/3748191.3748206}
\newcommand\vldbpages{3435 - 3448}
\newcommand\vldbvolume{18}
\newcommand\vldbissue{10}
\newcommand\vldbyear{2025}
\newcommand\vldbauthors{\authors}
\newcommand\vldbtitle{\shorttitle} 
\newcommand\vldbavailabilityurl{https://github.com/HPI-Information-Systems/P2E2-Erasure}
\newcommand\vldbpagestyle{empty} 

\begin{document}
\title{Meaningful Data Erasure in the Presence of Dependencies}

\author{Vishal Chakraborty}
\affiliation{%
  \institution{University of California (UC), Irvine}
}
\email{vchakrab@uci.edu}
\author{Youri Kaminsky}
\affiliation{%
  \institution{Hasso Plattner Institute,\\ University of Potsdam}
}
\email{youri.kaminsky@hpi.de}
\author{Sharad Mehrotra}
\affiliation{%
  \institution{UC, Irvine}
}
\email{sharad@ics.uci.edu}
\author{Felix Naumann}
\affiliation{%
  \institution{Hasso Plattner Institute,\\ University of Potsdam}
}
\email{felixnaumann@hpi.de}
\author{Faisal Nawab}
\affiliation{%
  \institution{UC, Irvine}
}
\email{nawabf@uci.edu}
\author{Primal Pappachan}
\affiliation{%
  \institution{Portland State University}
}
\email{primal@pdx.edu}
\author{Mohammad Sadoghi}
\affiliation{%
  \institution{UC, Davis}
}
\email{msadoghi@ucdavis.edu}
\author{Nalini Venkatasubramanian}
\affiliation{%
  \institution{UC, Irvine}
}
\email{nalini@uci.edu}

\begin{abstract}
Data regulations like GDPR require systems to support data erasure but leave the definition of "erasure" open to interpretation. This ambiguity makes compliance challenging, especially in databases where data dependencies can lead to erased data being inferred from remaining data. We formally define a precise notion of data erasure that ensures any inference about deleted data, through dependencies, remains bounded to what could have been inferred before its insertion. We design erasure mechanisms that enforce this guarantee at minimal cost. Additionally, we explore strategies to balance cost and throughput, batch multiple erasures, and proactively compute data retention times when possible. We demonstrate the practicality and scalability of our algorithms using both real and synthetic datasets.
\end{abstract}

\maketitle

\pagestyle{\vldbpagestyle}
\begingroup\small\noindent\raggedright\textbf{PVLDB Reference Format:}\\
\vldbauthors. \vldbtitle. PVLDB, \vldbvolume(\vldbissue): \vldbpages, \vldbyear.\\
\href{https://doi.org/\vldbdoi}{doi:\vldbdoi}
\endgroup
\begingroup
\renewcommand\thefootnote{}\footnote{\noindent
This work is licensed under the Creative Commons BY-NC-ND 4.0 International License. Visit \url{https://creativecommons.org/licenses/by-nc-nd/4.0/} to view a copy of this license. For any use beyond those covered by this license, obtain permission by emailing \href{mailto:info@vldb.org}{info@vldb.org}. Copyright is held by the owner/author(s). Publication rights licensed to the VLDB Endowment. \\
\raggedright Proceedings of the VLDB Endowment, Vol. \vldbvolume, No. \vldbissue\ %
ISSN 2150-8097. \\
\href{https://doi.org/\vldbdoi}{doi:\vldbdoi} \\
}\addtocounter{footnote}{-1}\endgroup

\ifdefempty{\vldbavailabilityurl}{}{
\begingroup\small\noindent\raggedright\textbf{PVLDB Artifact Availability:}\\
The source code, data, and/or other artifacts have been made available at \url{\vldbavailabilityurl}.
\endgroup
}

\section{Introduction}
\label{sec:intro}
Recently enacted data regulations~\cite{VDPA, CCPA, PIPEDA, GDPROffText, DataprotectGhana} have codified the right to erasure and established principles of data minimization and integrity. This has driven academic~\cite{shastri2019understanding, BuildingDelCompDB,sarkar2022constructing, sarkar:hal-01824058} and industry~\cite{CohnDeletionSocialNetwrks, fb_delete_nudges} efforts to address the challenges of compliant data erasure. In semantically rich databases, simply deleting the user-specified data is insufficient, as it may be inferred from remaining data. This problem arises in domains like social media, business, etc.---wherever applications store semantically linked user data.

Databases already support deletion of data beyond that specified by the user, particularly when required to preserve consistency~\cite{mysqlTriggers, gilad2020multiple}. For example, deleting a record in a parent table can trigger cascading deletions~\cite{UllmanBook} in child tables via foreign key constraints. Users can also define application-specific deletion logic through triggers~\cite{mysqlTriggers}. Cascading deletions—motivated by regulatory compliance—have been studied in~\cite{sarkar2022query, shastri2019understanding}, including ``shallow'' and ``deep'' deletions in graph databases~\cite{CohnDeletionSocialNetwrks} and annotation-driven deletions in~\cite{albab2023k9db}. However, these solutions are often bespoke to specific settings, and ad hoc—lacking formal guarantees. Critically, they adopt an operational view of deletion without grounding it in a principled notion of the user’s right to erasure, especially when semantic dependencies can enable inferences about deleted data.

Our goals in this paper are threefold: (a)~to define a principled notion of deletion for databases that prevents deleted data  from being inferred (by exploiting semantically dependent data) after deletion, 
(b)~to  develop effective and efficient ways to implement the developed deletion notion that  minimizes additional data to be deleted, and
(c)~to  understand (through a detailed experimental study across several domains and data sets) the implication and practical viability of such an approach.

\noindent\textbf{Pre-insertion Post Erasure Equivalence.}  We formally define \emph{Pre-insertion Post-Erasure Equivalence} (P2E2), a deletion principle that restricts inferences about deleted data using dependencies in the database to only those that were possible at the time of its insertion. P2E2 is defined at the cell level, with each cell assigned an expiration time by which it must be deleted. While some data regulations are vague on inference of deleted data using dependencies~\cite{chakraborty2023data}, some regulations~\cite{DataprotectGhana} require to prevent "reconstruction (of deleted data) in an intelligible form." Our formalization of data deletion in presence of data dependencies aligns with regulatory expectations for safe and effective deletion, selective retention, automated erasure and principles of data minimization and purpose limitation as outlined in CCPA~\cite{CCPA} \textsection 1798.105(d), PIPEDA~\cite{PIPEDA} \textsc{Principle} 4.5.5, GDPR~\cite{GDPROffText} \textsc{Art.} 5(1) \& 17, LGPD~\cite{LGPD} \textsc{Art.} 15, HIPAA~\cite{HIPPAA} 45 \textsc{CFR} \textsection 164.310(d)(2)(i), and PDPA~\cite{PDPA} \textsc{Sec.} 25.
\begin{example}
\label{exam:intro}
Consider a relation \textbf{R(Name, City, AreaCode)} with the constraint: \emph{if} \texttt{City = Irvine}, \emph{then} $\texttt{AreaCode} \in \{714, 949\}$ (implemented in PostgreSQL as a check constraint~\footnote{https://www.postgresql.org/docs/current/ddl-constraints.html}). Starting with the tuple \texttt{(Alice, NULL, NULL)}, first insert \texttt{AreaCode = 949}, resulting in the tuple \texttt{(Alice, NULL, 949)},  followed by \texttt{City = Irvine} results in \texttt{(Alice, Irvine, 949)}. Deleting \texttt{AreaCode} yields \texttt{(Alice, Irvine, NULL)}, from which one can now infer that \texttt{AreaCode} was in $\{714, 949\}$—information that was \emph{not inferable} at the time of \texttt{AreaCode}'s insertion. To satisfy \textsc{P2E2}, we must also delete \texttt{City}, ensuring no new inferences arise post-deletion. In contrast, if \texttt{City = Irvine} is inserted \emph{before} \texttt{AreaCode = 949}, the same deletion of \texttt{AreaCode} does not require deleting \texttt{City}, since the constraint was already active and the inference about \texttt{AreaCode} was possible prior to its insertion. This highlights the subtlety of \textsc{P2E2}: deletion must prevent \emph{new} inferences based on post-insertion context, which depends on the sequence of actions.
\end{example}

\noindent\textbf{Relational Dependency Rules.} To  formally define P2E2, we introduce \emph{relational dependency rules} (RDRs) that provide a simple, yet general, framework to express a variety of dependencies in data. 
RDRs use a SQL-like language and can express traditional dependencies including classes of both hard and soft constraints\footnote{Hard constraints are those that always hold while, soft constraints are likely to hold~\cite{chomicki2005minimal} on the database instance, though are 
not guaranteed to hold.}, such as functional \& inclusion dependencies, denial constraints \cite{lopatenko2007complexity,bertossi2011database} and correlation constraints, respectively. RDRs, in addition, allow constraints  to be limited  to only a subset of data that satisfies the SQL query enabling specifications of conditional constraints such as (conditional) functional dependencies \cite{fan2008conditional,  disc_corr, bleifuss2024discovering, bohannon2007conditional}, and similarity inclusion dependencies~\cite{kaminsky2023discovering}.
Frameworks, similar to RDRs, to express semantic constraints in database have  previously been  proposed in a 
variety of data processing contexts ranging from database design and cleaning~\cite{nadeef, rekatsinas2017holoclean}, consistent query answering \cite{arenas1999consistent}, to database repair \cite{bertossi2011database}. 
We adopt RDRs as they are based on SQL and hence expressive, and intuitive.
Additionally, RDRs can allow specification of aggregation constraints (often found in organizational databases) which  existing framework for specifying constraints such as~\cite{gilad2020multiple} do not consider. Furthermore, RDRs can express regulatory-motivated semantic annotations~\cite{albab2023k9db, CohnDeletionSocialNetwrks}, dependencies discovered from data~\cite{GDPRizer}, and fine-grained erasure logic—e.g., Meta’s rule to delete user comments but retain messages~\cite{FBDelAcc}, or selective retention under GDPR Art. 17(3)~\cite{GDPROffText}.

\noindent \textbf{Minimizing overhead.} Given a set of RDRs\footnote{We assume a list of RDRs capturing semantic constraints in the data as input. 
}, 
and a data item to be deleted, additional data may need to be deleted to ensure P2E2.  Such additional deletions, as illustrated earlier, depend on the database state, both at the time of
deletion, as well as the time of insertion of the data item being deleted. In Example~\ref{exam:intro}, the database state  at the deletion time of $\texttt{AreaCode}$ was \texttt{(Alice, Irvine, 465)}. However, depending upon the state at the time of \emph{insertion} (viz. \texttt{(Alice, NULL, NULL)} versus \texttt{(Alice, Irvine, \textit{NULL})}) the set of 
additional deletions required to implement P2E2 differed. Realizing P2E2  requires  tracking what changes  were made to the  database  between the time the data was inserted to the time it is to be deleted, and to develop a logic to reason with RDRs  to determine if such changes can cause   leakage beyond what was already possible prior to insertion.  In addition, to minimize overhead, 
 we would further like the set of additional deletions required (to implement P2E2) to be minimal.
We note that such a logic to realize P2E2 cannot be easily encoded as simple
 deletion rules in the form of triggers\footnote{Trigger conditions only check the database state before and after the triggering event and, as such,  
 are independent of the sequence of events that led to the state.}. We thus develop algorithms to realize P2E2 with minimal deletions and realize them in a middleware layer on top of the database.

Our problem of computing a minimal set of deletions to satisfy P2E2, formalized as the $\textsc{Opt-P2E2}$ problem, is related to the well-studied problem of minimal repair~\cite{arenas1999consistent, chomicki2005minimal}. The minimal repair problem takes as input a database $D$ that is inconsistent with respect to a set of constraints $C$ (e.g., denial constraints~\cite{chomicki2005minimal, lopatenko2007complexity, staworko2007declarative}) and computes a minimal set of modifications that restore consistency.

In contrast, P2E2 starts with a database consistent with respect to its constraints and a data item $c$ to be deleted. The goal is to find a minimal subset of dependent data whose deletion ensures that no new inferences about $c$ can be made beyond what was inferable at the time of its insertion. While both problems aim to minimize deletions, their objectives—and hence their solutions—differ. Repair algorithms cannot be directly applied to enforce P2E2. Therefore, we leverage RDRs to model data dependencies and design erasure mechanisms tailored to ensure P2E2.

\noindent \textbf{Erasure mechanisms.} 
We develop mechanisms to enforce P2E2 for: 
(a)~\emph{demand-driven erasure}, where users request deletion at any time, and 
(b)~\emph{retention-driven erasure}, where data is deleted after a predefined retention period~\cite{BuildingDelCompDB}, as in WhatsApp’s disappearing messages feature~\cite{WhatsAppDis}. In both cases, the goal is to identify a minimal set of dependent data to delete to satisfy P2E2.
We propose multiple strategies to compute this minimal set, including ILP, bottom-up tree traversal, top-down traversal (trading precision for throughput), and batched erasure. These methods offer different trade-offs in terms of overhead (time and space) and deletion scope. To further reduce overhead, we tailor optimizations to each erasure type. For demand-driven erasure, we introduce a grace period to batch deletions. For retention-driven erasure, we leverage advance knowledge of deletion times to schedule erasures cost-effectively—especially when derived data (e.g., aggregates or materialized views) must be reconstructed after base data is deleted.

\noindent \textbf{Evaluation.} We evaluate our approaches on real and synthetic datasets under various workloads, analyzing the cost and performance impact of P2E2. We compare exact and approximate algorithms, studying trade-offs between computational overhead and cost. Additionally, we examine the effects of batching, varying grace periods, and pre-computing retention periods for derived data. 
Our evaluation on five data sets shows that, on average, the number of extra deletions to guarantee P2E2 for a given cell is low and depends on factors such as the number of cells in a tuple, dependencies, and the number of insertions and deletions. 

\begin{figure*}[t]
    \centering
    \begin{subfigure}{.9\linewidth}
         \includegraphics[trim={0mm 205mm 0mm 0mm}, clip, width=\linewidth]{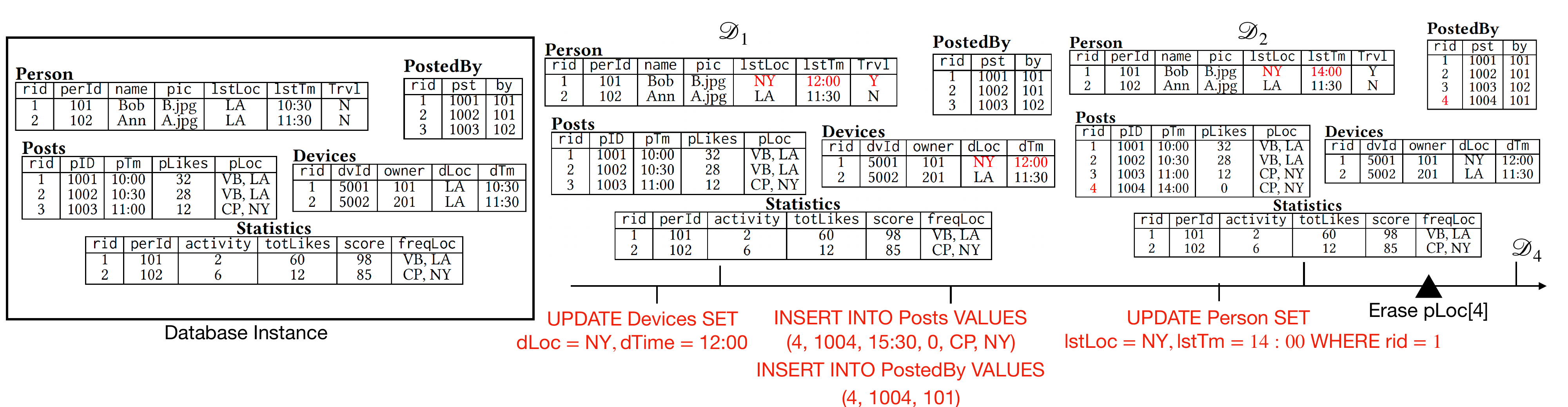}
    \caption{Database Instance in a social media platform and timeline of operations for running example.}
    \label{fig: timeline}
    \end{subfigure}
    \hfill
    \begin{subfigure}{.4\linewidth}
    \small
        \begin{tabular}{@{}ll@{}}
            & \\
            Dependence:& $\schele{totLikes}(R) \dep \schele{pstLikes}(M)$ \\ 
            Condition:& $\text{SELECT } \schele{S.rid}, \schele{P.rid} \text{ AS } R, M$ \\ 
            &$\text{FROM } \schele{Statistics} \text{ } \schele{S}, \schele{Posts} \text{ } \schele{P}, \schele{Person} \ \schele{I}, \schele{PostedBy} \ \schele{B}$\\
            &$\text{WHERE} \schele{S.perID} = \schele{I.perID}\text{ AND } \schele{B.pst} = \schele{P.pID} $\\
            & $\text{ AND } \schele{B.by} = \schele{I.perID}$ \\
        \end{tabular}
        \caption{RDR R1}
        \label{fig:R1}
    \end{subfigure}
    \hfill
     \begin{subfigure}{.5\linewidth}
     \small
        \begin{tabular}{@{}ll@{}}
            Dependence: &  $\schele{lastLoc}(U) \dep \schele{pLoc}(M)$  \\ 
            Condition: & $\text{SELECT } \schele{I.rid}, \schele{P.rid} \text{ AS } U, M$ \\
                      &  $\text{FROM } \schele{Person I}, \schele{Devices D}, \schele{PostedBy B}, \schele{Posts P} \text{ (}$\\ 
                      & $\text{SELECT } \schele{B.by} \text{ AS } \schele{by}, \text{ MAX(} \schele{P.pTM} \text{) AS } \schele{max\_pTm} \text{ FROM}$\\
                      &$\schele{Postedby B}, \schele{Posts P} \text{ WHERE } \schele{ B.pst}= \schele{P.pID}$\\ &$\text{ GROUP BY } \schele{B.by} \text{ ) AS } \schele{last\_posts} \text{ WHERE }$ \\
                      &$\quad \schele{I.perID = D.owner} \text{ AND } \schele{I.perID = B.by} \text{ AND } $\\
                      &$\quad \schele{B.by = last\_posts.by} \text{ AND }$ \\
                      & $\schele{P.pTm = last\_posts.max\_pTM} \text{ AND } \schele{P.pTm > D.dTm} $ \\
        \end{tabular}
        \caption{RDR R2}
        \label{fig:R2}
    \end{subfigure}
    \Description{Database instance, database states, and data dependencies (RDRs) for running example.}
    \caption{Database instance, database states, and data dependencies (RDRs) for running example.}
\end{figure*}
Section (Sec.)~\ref{sec: form_data_erase} formalizes data erasure and introduces its semantic guarantee, P2E2. Sec.~\ref{sec: RDR} presents Relational Dependency Rules, and technical results for reasoning about cell-level erasure. Sec.~\ref{sec: demand-driven} develops algorithms for demand-driven erasure, while Sec.~\ref{sec: retention} addresses retention-driven deletion. We evaluate our approaches and analyze P2E2's overheads in Sec.~\ref{sec: eval}. Sec.~\ref{sec: related_works} reviews related work, and Sec.~\ref{sec: conclusion} concludes with directions for future research.

\section{Preliminaries}
\label{sec: form_data_erase}
\begin{table}[b]
    \centering
    \small
    \begin{tabular}{ll}
    \toprule
    Symbol & Meaning \\
    \midrule
    $\mathcal{S}=(\mathcal{R}, \mathcal{A})$ & Schema with sets $\mathcal{R}$ of relations \& $\mathcal{A}$ of attributes  \\
     $\mathbf{R}_i$, $\texttt{A}_j$ & A relation $\mathbf{R}_i \in \mathcal{R}$ and attribute $\texttt{A}_j \in \mathcal{A}$ \\
    $rid_k$ & Record-ID uniquely identifying records\\
     $\mathcal{D}_t$ & Database state, i.e., set of cells in database at time $t$ \\
    $Cells(\cdot)$ & Operator returning all cells in the given argument \\
    $A_j(rid_k)$ & Cell containing the value of attribute $A_j$ of $rid_k$ \\
    $\kappa(c)$ & Creation (insertion) timestamp of cell $c$ \\
    $\eta(c)$ & Expiration timestamp of cell $c$ (erasure time) \\
    $Val(\texttt{A}(\boldsymbol{x}), \mathcal{D}_t)$ & Value of attribute function $\texttt{A}(\boldsymbol{x})$ in $\mathcal{D}_t$ \\
    $dep(\texttt{A}(\boldsymbol{x}) \mid \mathcal{D}_t)$ & Set of dependencies on $\texttt{A}(\boldsymbol{x})$ in $\mathcal{D}_t$ \\
    $\delta^-$ & Instantiated relational dependency rule (RDR) \\
    $\Delta^- (\mathcal{D}_t)$ & Set of all instantiated RDRs in $\mathcal{D}_t$ \\
    $Head(\delta^-)$ & Head of instantiated RDR $\delta^-$ \\
    $Tail(\delta^-)$ & Tail of instantiated RDR $\delta^-$ \\
    \bottomrule
    \end{tabular}
    \caption{Notation Table}
    \label{tab:notation}
\end{table}
In this section, we introduce the formal semantics of data erasure and adopt standard notation and concepts from \cite{CausalRelationalLearning,geiger1991axioms}. But first we introduce a running example, which we use throughout the paper to illustrate notations and explain the intuition behind P2E2.

\noindent \textbf{Running Example.} Consider a social media platform where users post content, upload photos, and use location-based services. The platform maintains the following database tables (Fig.~\ref{fig: timeline}):  

\noindent
- \schele{Person}: Stores user details, including name, profile picture, travel status (\schele{Trvl}), and last known location (\schele{lstLoc}).  

\noindent
- \schele{Posts}: Records user posts with timestamps, locations, and like counts (\schele{pLikes}) and \schele{PostedBy}: Tracks the author of each post.  

\noindent
- \schele{Device}: Stores the last known location (\schele{dLoc}) of a user’s device.  

Attribute \schele{lstLoc} in \schele{Person} updates when the user’s device location (\schele{dLoc}) changes or makes a new post with location (\schele{pLoc}).  

\noindent
- The platform also maintains a \schele{Statistics} table for analytics, tracking average activity (\schele{activity}), most frequented location (\schele{freqLoc}), and total likes (\schele{totLikes}). The system enforces:
(1) Updates to \schele{pLikes} propagate to \schele{totLikes} for the user;
(2) \schele{Trvl} in \schele{Person} is set to \schele{Y} if \schele{freqLoc} $\neq$ \schele{lstLoc};
(3) Other \schele{Statistics} attributes (e.g., \schele{activity}) update periodically (e.g., weekly).


In the following, we present some notation and discuss our data and retention model required to formalize the problem setting. 

\noindent \textbf{Functional Data Model. } We extend the functional data model~\cite{shipman1981functional} and its recent adaptation~\cite{CausalRelationalLearning}. A database $\mathcal{D}$ consists of relations $\mathcal{R} = \{\boldsymbol{R}_1, \ldots, \boldsymbol{R}_m \}$, each with attributes $\boldsymbol{R}_i^{\textit{attr}} = \{\texttt{A}_1, \ldots, \texttt{A}_{n_j}\}$. Attributes are referred to as $\texttt{A}_j$ when the relation is clear from context. In Fig.~\ref{fig: timeline}, \schele{Person} is a relation, attributes \schele{perId}, \schele{name}, \schele{profPic}, and \schele{lastLoc}. Each relation $\boldsymbol{R}_i$ contains records $r_k^i$, uniquely identified by $\schele{rid}_k$, its record-Id (\schele{rid}). Records consist of cells, each corresponding to an attribute. We use the operator $\textit{Cells}(\cdot)$ which returns all the cells in the argument.
To an attribute $\texttt{A}_j \in \boldsymbol{R}_i^{\textit{attr}}$, we associate the attribute function $\texttt{A}_j: \boldsymbol{R}_i(\schele{rid}) \rightarrow \textit{Cells}(\mathcal{D})$ that takes in as input an \schele{rid} of a relation and returns the cell corresponding to attribute $\texttt{}{A_j}$. A cell is denoted as $\texttt{A}_j(\boldsymbol{R}_i(\schele{rid}_k))$, or $\texttt{A}_j(\schele{rid}_k)$ when the relation is clear. In the database instance in Fig.~\ref{fig: timeline}, \schele{name}(\schele{1}) refers to the cell containing ``Bob''. Each cell has an associated deletion cost given by the function $\textit{Cost}: \mathcal{D} \rightarrow \mathbb{R}$. A relational functional schema is $\mathcal{S} = (\mathcal{R}, \mathcal{A})$ where $\mathcal{A} = \bigcup_{i=1}^m \boldsymbol{R}_i^{\textit{attr}}.$

\noindent
\textbf{Database State.} At any time $t$, we write $\mathcal{D}_t$ to denote the \emph{database state}, i.e., the set of all records in the database at time $t$ and their cell values. The value of a cell $c$ in $\mathcal{D}_t$ is given by $\textit{Val}(c, \mathcal{D}_t) \in \textit{Dom}(\texttt{A}_j)$ where $\textit{Dom}(\texttt{A}_j),$ the domain of $\texttt{A}_j,$ is the set of all possible values $\texttt{A}_j$ can take including $\textit{NULL}.$ As an example, in Fig.~\ref{fig: timeline}, $\textit{Val}(\schele{name}(\schele{1}), \mathcal{D}_1) = \schele{Bob}.$  With $\mathcal{D}_{t^-}$ and $\mathcal{D}_{t^+}$ we denote the states immediately before and after $\mathcal{D}_t$, respectively. A cell's value becomes \textit{NULL} (empty) when it is erased, and erasure of a record entails erasure of all cells within it which are not empty.

\noindent
\textbf{Types of Data.}
Relations are classified as \emph{base} or \emph{derived}. Base relations store personal data on which the user has direct control (request deletion, rectification, etc.). Derived relations contain data which result from processing base and/or other derived relations. In the example, \schele{Posts} is a base relation, while \schele{Statistics} is derived.

\textit{Base Relations.}
A cell $c$ in base relations has: (1)~creation timestamp $\kappa(c)$ \textemdash\ the time at which it is inserted, (2)~erasure/expiration timestamp $\eta(c)$ \textemdash\ the time at which it is deleted. The retention period of a cell is the interval between $\kappa(c)$ and $\eta(c)$. When a record is inserted, for each cell $c$ in the record, $\kappa(c)$ is initialized with the time at which the record was inserted. The expiration time of each cell is set to a fixed or user-specified time at which it needs to be deleted from the database. For a base cell $c$, users can adjust $\eta(c)$ to enable demand-driven erasure, which overrides the default retention period. Updating a cell is treated as an erasure followed by insertion, updating $\kappa(c)$ to the time of the update but retaining the original $\eta(c)$. Base data erasure satisfies P2E2.

\textit{Derived Relations.}
Derived data are computed from base or other derived relations, with periodic recomputation. For a derived cell $c$, we denote with $\textit{freq}(c)$ the time period in which $c$ has to reconstructed at least once. For example, $\textit{freq}(c)$ for $\schele{score}$ is 30 days, and for $\schele{totLikes}$, 7 days. When a derived cell $c$ is recomputed, its creation timestamp $\kappa(c)$ is updated to the recomputation time, while $\textit{freq}(c)$ remains unchanged. Derived data lack explicit erasure timestamps, as they may be reconstructed after base data erasure to prevent inferences. Derived data are deleted only when no longer required, ceasing further reconstruction. Particularly, users cannot directly ask derived data to be deleted.

\section{P2E2}
\label{sec: RDR}
In a database, the value of a cell often depends on that of other cells. In this section, we will formally define RDRs as a means 
of expressing data dependencies. We discuss how RDRs can be instantiated, and how we can reason about inferences using RDRs.
 
\subsection{Specifying Background Knowledge by RDRs}
\label{sub: RDRs}
RDRs express dependency (background knowledge) among cells without necessarily specifying how the cells are dependent.
An RDR consists of two parts \textemdash\ a dependence statement (which itself has two parts:
a head and tail) and a condition. The dependence statement specifies the dependency among cells, while
the condition, is a query that identifies the cells on which a dependency holds and returns the \schele{rid}s corresponding to those cells.

\begin{definition}[Relational dependency Rules]
\label{defn: rules}
     Let $\boldsymbol{S}=$$(\mathcal{R}, \mathcal{A})$ be a relational functional schema where $\texttt{A}, \texttt{A}_{\texttt{1}}, \ldots, \texttt{A}_{\texttt{q}} \in \mathcal{A}$  are attribute functions, and $Q$ is a SQL query over the schema $\boldsymbol{S}$ that returns the \schele{rid}s $\boldsymbol{X}, \boldsymbol{X}_1, \ldots, \boldsymbol{X}_p$ of records that satisfy the condition of the query. A \emph{relational dependency rule (RDR)} is given by
 \begin{align}
   \label{eqn: detRDR_defn}
        \text{Dependence: }&\texttt{A}(\boldsymbol{X}) \dep \texttt{A}_\texttt{1}(\boldsymbol{X}_1), \ldots, \texttt{A}_{\texttt{p}}(\boldsymbol{X}_p) \qquad
        \text{Condition: } Q
   \end{align}
\end{definition}
 In Eqn.~\ref{eqn: detRDR_defn}, the \emph{condition} $Q$ 
 identifies  \schele{rid}s of a subset of records (from the set of relations in $\mathcal{R}$) such that the
 records contain the 
 attributes $\texttt{A}, \texttt{A}_1, \ldots, \texttt{A}_p$ which are dependent.  The dependence part of the RDR  $\texttt{A}(\boldsymbol{X}) \dep \texttt{A}_\texttt{1}(\boldsymbol{X}_1), \ldots, \texttt{A}_{\texttt{p}}(\boldsymbol{X}_p) $ uses attribute function notation to express
the dependency between attribute values amongst the selected records. 
In particular, the RDR expresses that value the attribute $\texttt{A}(\boldsymbol{X})$ (\emph{head} of the RDR) can take is {\bf  not independent} of the value of the attributes $\texttt{A}_\texttt{1}(\boldsymbol{X}_1), \ldots, \texttt{A}_{\texttt{p}}(\boldsymbol{X}_p)$ (\emph{tail} of the RDR). Thus, instantiated values of the attributes in the tail of the RDR can enable inference about the value of the head of the attribute. 

We illustrate the RDR notation using a couple of semantic dependencies among the data in our example. In Fig.~\ref{fig:R1}, RDR R1 states the dependency between the number of likes (\schele{plikes} in \schele{Posts} table) for a post authored by a person u and the total likes (\schele{totlikes} in \schele{Statistics} table) of u.
The condition of R1 chooses \schele{rid} pairs corresponding to the \schele{rid}s of the records in the \schele{Statistics} and \texttt{Post} tables s.t.\ the two records  are for the same
person (due to join conditions in the WHERE clause). 

As another illustration, we consider R2 (Fig.~\ref{fig:R2}) that states that the last location \schele{lstLoc} of a person u and the location of the latest post (\schele{pLoc}) made by u are dependent if the post was made before the time the location of the device of u was collected.  
We present a few other dependencies for our running example which will be used later. Let u be a person with a device d and makes a post p:
\begin{itemize}  
    \item {\bf R3}: $\schele{lstLoc} \dep \schele{dLoc},$ i.e., the last known location (\schele{lstLoc}) of u depends on the location of d (\schele{dLoc} collected at $\schele{dTm}$) if the last post made by u, \schele{pTm} is earlier than \schele{dTm}.

    \item {\bf R4}: The location of p (\schele{pLoc}) and \(u\)'s most frequented location (\schele{freqLoc}) are dependent, i.e., $\schele{freqLoc}(R) \dep \schele{pLoc}(M).$
    
    \item {\bf R5}: The location of d $\schele{dLoc} \dep \schele{pLoc}$ of p if d's location (\schele{dLoc}) was collected within an hour of the time of p.

    \item {\bf R6}: For u, $\schele{Trvl} \dep \schele{freqLoc}, \schele{lstLoc}$. 
\end{itemize}
 
RDRs may be discovered using existing work (such as~\cite{GDPRizer, albab2023k9db} which, motivated by data regulations, discovers data dependencies and annotations, and others like~\cite{bleifuss2024discovering, kaminsky2023discovering}) or through data analysis. In our evaluations, we derived dependencies using existing work and manually. Moreover, rules used for data repairs, data cleaning, as well as provenance annotations can be easily expressed as RDRs.

\noindent \textbf{Instantiated RDRs.}
To use RDRs for specific cells, we need to define \emph{instantiated RDRs}. Instantiations of a RDR, on a database state $\mathcal{D}_t,$ are all possible dependencies (in $\mathcal{D}_t$) between the relevant attribute functions (in $\mathcal{D}_t$) which are specified in the dependence statement of the RDR and satisfy the condition of the RDR. When an attribute function is in the dependence as well as the condition, it is dropped from the latter.  An instantiated RDR comprises just the dependence statement of the corresponding RDR, i.e., the head and the tail of the RDR, with all the variables $\boldsymbol{X}, \boldsymbol{X}_1, \ldots, \boldsymbol{X}_p$ substituted with constants from the database state $\mathcal{D}_t$ which are returned by the condition of the RDR. For example, the instantiations of RDR $R1$ in Fig.~(\ref{fig:R1}) for $\schele{Person}(\schele{1})$ are:  $\delta^-_2: \schele{totLikes}(\schele{1}) \dep 
     \schele{pstLikes}(\schele{1})$,
     $\delta^-_3: \schele{totLikes}(\schele{1}) \dep \schele{pstLikes}(\schele{2})$, and 
     $\delta^-_4: \schele{totLikes}(\schele{1}) \dep \schele{pstLikes}(\schele{4})$ in $\mathcal{D}_2.$ 

\begin{definition}[Instantiations of RDRs]
Let $\mathcal{D}_i$ be an instance over $\boldsymbol{S} = (\mathcal{R}, \mathcal{A})$. Given an RDR as in Eqn.~\ref{eqn: detRDR_defn},
for an instantiated attribute function $\texttt{A}_k(\boldsymbol{x}_k),$ we associate the \emph{instantiated} RDR, denoted $\delta^-(\mathcal{D}_i, \texttt{A}_k(\boldsymbol{x}_k))$, obtained by assigning to the variables $\boldsymbol{X} = \boldsymbol{x}, \boldsymbol{X}_1 = \boldsymbol{x}_1, \ldots, \boldsymbol{X}_p = \boldsymbol{x}_p$ such that  $\textit{Val}(\texttt{A}_1(\boldsymbol{x}_1), \mathcal{D}_i),\ldots,\textit{Val}(\texttt{A}_p(\boldsymbol{x}_p), \mathcal{D}_i)$ are returned by the query $Q$ evaluated on $\mathcal{D}_i$. 
When clear from the context, we drop the database state $\mathcal{D}_i$ and $\texttt{A}_k (\boldsymbol{x}_k)$ from $\delta^-(\mathcal{D}_i, \texttt{A}_k(\boldsymbol{x}_k))$ and write $\delta^-.$ An instantiated RDR is of the following form where $\texttt{A}_i(\boldsymbol{x}_i)$ is the instantiated attribute function for $\texttt{A}_i(\boldsymbol{X}_i).$ 
\begin{align}
\label{eqn: instantiated_RDR}
    \delta^-: \texttt{A}(\boldsymbol{x}) \dep
        \texttt{A}_{1}(\boldsymbol{x}_{1}), \ldots, \texttt{A}_p(\boldsymbol{x}_p)
\end{align}
\end{definition}

We denote a set of RDRs with $\Delta^-.$
Given a database state $\mathcal{D}_t$, we denote with $\Delta^-(\mathcal{D}_t)$ the set of all instantiated RDRs on that state. With $\textit{Head}(\delta^-)$, we refer to $\{\texttt{A}(\boldsymbol{x})\},$ the head of the instantiated RDR. The tail, denoted $\textit{Tail}(\delta^-)$  is given by the set $\{\texttt{A}_1(\boldsymbol{x}_1), \ldots, \texttt{A}_p(\boldsymbol{x}_p)\}$. The condition is dropped. We denote with $\textit{Cells}(\delta^-)$ the set of all the attribute functions in $\delta^-$, i.e., 
$\textit{Head}(\delta^-) \cup \textit{Tail}(\delta^-).$


\subsection{Dependency Sets}
Given a set of instantiated RDRs, we define the notion of dependency sets. Dependency sets capture how an attribute function $\texttt{A}(\boldsymbol{x})$ can be probabilistically influenced by, or can influence other data in the database. Instantiated RDRs can lead to the inference of an attribute function $\texttt{A}(\boldsymbol{x})$ in two ways: (1)~direct and (2)~indirect.

\noindent
 \textbf{Direct inference.} Direct inference takes place through explicitly stated dependencies that involve $\texttt{A}(\boldsymbol{x}).$ 
\begin{example}
\label{ex: reasoning}
The instantiation $\delta^-_5: \texttt{freqLoc}(\schele{1}) \dep \texttt{pLoc}(\schele{4})$ of R4 and the instantiation $\delta^-_6: \schele{lastLoc}(\schele{1})  \dep \texttt{pLoc}(\schele{4})$ of R2 lead to the direct inference of $\schele{pLoc}(\schele{4}).$
\end{example}

\noindent\textbf{Indirect Inference.} An indirect inference on $\texttt{A}(\boldsymbol{x})$ exists when a sequence of instantiated RDR can be used to infer $\texttt{A}(\boldsymbol{x})$ through shared elements between each pair of instantiated RDRs. Continuing Eg.~\ref{ex: reasoning}, let $\delta^-_7: \schele{lastLoc}(\schele{1}) \dep \schele{dLoc}(\schele{1})$ be another instantiated RDR. Observe that $\schele{pLoc}(\schele{1})$ does not directly depend on $\schele{devLoc}(\schele{1}).$ However, $\schele{pLoc}(\schele{1})$ depends on $\schele{lasLoc}(\schele{1})$ through $\delta^-_6$, which in turn depends on $\schele{dLoc}(\schele{1})$.

\begin{definition}
    Given a database state $\mathcal{D}_t$ and an attribute function $\texttt{A}(\boldsymbol{x}) \in \mathcal{D}_t,$ and a set of $\Delta^-(\mathcal{D}_t)$ of instantiated RDRs, we define the set of dependencies on $\texttt{A}(\boldsymbol{x})$ in $\mathcal{D}_t,$ denoted $\textit{dep}(\texttt{A}(\boldsymbol{x}) \mid \mathcal{D}_t),$ to contain $\delta^-_i \in \Delta^-(\mathcal{D}_t)$ such that for all cells $c \in \textit{Cells}(\delta^-_i),$ we have $\textit{Val}(c \mid \mathcal{D}_t) \neq \textit{NULL}$ and one of the following holds:(1) $\texttt{A}(\boldsymbol{x}) \in \textit{Cells}(\delta^-_i);$ (2) there exists $\delta^-_1, \ldots, \delta^-_{\ell}, \ldots, \delta^-_K$ in $\Delta^-(\mathcal{D}_t)$ such that, for $1 <i \leq K,$ we have $\textit{Tail}(\delta^-_{\ell}) \cap \textit{Head}(\delta^-_{\ell +1}) \neq \emptyset$ and $\texttt{A}(\boldsymbol{x}) \in \textit{Cells}(\delta^-_1).$
\end{definition}
\subsection{Data Erasure Semantics}
\label{sub: del_defn}
We formalize the semantic guarantees of data erasure in this section. We assume that at a given time, the database owner has access to the entire database at that time, and the dependencies. We do not consider adversarial scenarios wherein a malicious database owner maintains a copy of data secretly. Incorporating such a possibility goes well beyond the scope of erasure we are considering here.

Note that when a cell $\texttt{A}(\boldsymbol{x})$ expires at $t_e$, its value is set to $\textit{NULL}.$ We write $\textit{Val}(\texttt{A}(\boldsymbol{x})) \leftarrow \textit{val}$ to denote the assignment of the value  $\textit{val}$ to the cell $\texttt{A}(\boldsymbol{x}).$ In particular, with $\mathcal{D}_{t_e^+} \cup \{ \textit{Val}(\texttt{A}(\boldsymbol{x})) \leftarrow \textit{val} \}$ we denote the state that is identical to $\mathcal{D}_{t_e^+}$ except that the cell $\texttt{A}(\boldsymbol{x})$ which has the value $\texttt{val}.$
We now formally define P2E2.
\begin{definition}[Pre-insertion Post Erasure Equivalence (P2E2)]
\label{defn: P2E2}
     Given a set $\Delta^-$ of RDRs, a cell $\texttt{A}(\boldsymbol{x})$ with insertion time $\kappa(\texttt{A}(\boldsymbol{x})) = t_b$, expiration time $\eta(\texttt{A}(\boldsymbol{x})) = t_e$ and
     $\textit{Val}(\texttt{A}(\boldsymbol{x}), \mathcal{D}_{t_e}) = val$ that is not \textit{NULL}, we say that \emph{pre-insertion post erasure equivalence (P2E2)} holds for $\texttt{A}(\boldsymbol{x})$ if: $ \textit{dep}(\texttt{A}(\boldsymbol{x}) \mid \mathcal{D}_{t_e^+}) \cup \{\textit{Val}(\texttt{A}(\boldsymbol{x})) \leftarrow \textit{val} \})) \subseteq \textit{dep}(\texttt{A}(\boldsymbol{x}) \mid \mathcal{D}_{t_b}) 
    .$
\end{definition}
Informally, P2E2 states that the set of dependencies on an attribute function when it is inserted is the same as the set of dependencies on the attribute function after it has expired. Note that $\texttt{A}(\boldsymbol{x})$ must be set to $\textit{NULL}$ after it has expired. In Fig.~\ref{fig: timeline}, suppose Bob wants the location of their latest post (i.e., attribute \schele{pLoc} in table \schele{Posts}, \texttt{rid}=4 with \texttt{pID}=1004) to be forgotten. We say P2E2 guarantee holds for \schele{pLoc(4)} if the dependencies on \schele{pLoc(4)} in state $\mathcal{D}_1$ is the same as that in the state after the $\blacktriangle,$ i.e., state $\mathcal{D}_4$.

We can now formally define the problem of minimal erasure.
\begin{definition}[\textsc{Opt-P2E2}]
Given a database state $\mathcal{D}_t$, the set of instantiated RDRs $\Delta^-(\mathcal{D}_t),$ and a cell $\texttt{A}(\boldsymbol{x})$ in $\mathcal{D}_t.$  Find a set $\mathcal{T} = \{ \texttt{A}(\boldsymbol{x})\} \cup \{\texttt{A}_i(\boldsymbol{x}_i) \mid 1 \leq i \leq |\mathcal{T}|\}$ such that when the value of each $\texttt{A}_i(\boldsymbol{x}_i)$ is set to $\textit{NULL},$ P2E2 holds in $\mathcal{D}_{t^+}$ for $\texttt{A}(\boldsymbol{x})$ and  $\sum_{\texttt{A}_i(\boldsymbol{x}_i) \in \mathcal{T}} \textit{Cost}(\texttt{A}_i(\boldsymbol{x}_i))$ is minimized.
\end{definition}
\begin{theorem}
\label{thm: hardness}
    The \textsc{Opt-P2E2} problem is \textsc{NP-Hard}.
\end{theorem}
\subsection{Identifying Relevant Dependencies}
\label{sub:id_RDR}
Given a set of RDRs, and an attribute function $\texttt{A}(\boldsymbol{x})$ for which P2E2 has to be guaranteed, where $\kappa(\texttt{A}(\boldsymbol{x})) = t_b$ and $\eta(\kappa)(\boldsymbol{x}) = t_e,$ we need to instantiate RDRs to determine the sets $\textit{dep}(\texttt{A}(\boldsymbol{x}) \mid \mathcal{D}_{t_b})$ and $\textit{dep}(\texttt{A}(\boldsymbol{x}) \mid \mathcal{D}_{t_e})$ to compute $\Delta^-(\textit{P2E2}, \texttt{A}(\boldsymbol{x}) ) =  \textit{dep}(\texttt{A}(\boldsymbol{x}) \mid \mathcal{D}_{t_e}) \setminus \textit{dep}(\texttt{A}(\boldsymbol{x}) \mid \mathcal{D}_{t_b}).$ Then we address each dependency in $\Delta^-(\textit{P2E2}, \texttt{A}(\boldsymbol{x}) )$ to ensure that P2E2 holds.

\noindent
\textbf{Constructing} $\textit{dep}(\texttt{A}(\boldsymbol{x}) \mid \mathcal{D}_t)$. Recall that inference using RDRs can be direct or indirect.  Direct inference occurs when an instantiated RDR $\delta^-$ contains $\texttt{A}(\boldsymbol{x}) \in \textit{Cells}(\delta^-).$ Therefore we need to instantiate all such RDRs where $\texttt{A}(\boldsymbol{X})$ is in the head or the tail of the dependence. To prevent indirect inference on $\texttt{A}(\boldsymbol{x}),$ we need to consider all RDRs that lead to the direct inference of some attribute function $\texttt{A}'(\boldsymbol{x})$ such that $\texttt{A}'(\boldsymbol{x})$ leads to to direct inference of $\texttt{A}(\boldsymbol{x}).$

With both sets of dependencies constructed as above, the difference $\Delta^-(\textit{P2E2}, \texttt{A}(\boldsymbol{x}) ) = \textit{dep}(\texttt{A}(\boldsymbol{x}) \mid \mathcal{D}_{t_e} ) \setminus \textit{dep}(\texttt{A}(\boldsymbol{x}) \mid \mathcal{D}_{t_b})$ can be readily identified.  However, constructing sets of dependencies is computationally intensive as it entails recursively instantiating RDRs to account for both direct and indirect inference.

\noindent\textbf{Constructing }$\Delta^-(\textit{P2E2}, \texttt{A}(\boldsymbol{x}) ).$ We note that instead of constructing the two sets of dependencies, namely, $\textit{dep}(\texttt{A}(\boldsymbol{x}) \mid \mathcal{D}_{t_e} )$ and $\textit{dep}(\texttt{A}(\boldsymbol{x}) \mid \mathcal{D}_{t_b}),$ and then computing their difference, we can directly construct $\Delta^-(\textit{P2E2}, \texttt{A}(\boldsymbol{x}) )$ To that end, let $E$ be the set of cells erased and $N$ be the set of cells inserted after state $\mathcal{D}_{t_b}.$ Therefore, $(\mathcal{D}_{t_b} - E) \cup N = \mathcal{D}_{t_e}.$  We first observe that the dependencies on $\texttt{A}(\boldsymbol{x})$ in $\mathcal{D}_{t_b} \setminus E$ are necessarily contained in the set of dependencies on $\texttt{A}(\boldsymbol{x})$ in $\mathcal{D}_{t_b}.$ This is because since some cells are erased, no new dependencies are introduced. Therefore, we only need to focus on the set of dependencies on $\texttt{A}(\boldsymbol{x})$ introduced by the cells in $N.$

Observe that not all dependencies in the set $\textit{dep}(\texttt{A}(\boldsymbol{x}) \mid \mathcal{D}_{t_e})$ are in the desired set $\Delta^-(\textit{P2E2}, \texttt{A}(\boldsymbol{x}))$ of dependencies that violate P2E2. More specifically, $\textit{dep}(\texttt{A}(\boldsymbol{x}) \mid \mathcal{D}_{t_e}),$ also contains dependencies that existed in the state $\mathcal{D}_{t_b} - E.$ Observe that if all the cells of an instantiated RDR $\delta^- \in \textit{dep}(\texttt{A}(\boldsymbol{x}) \mid \mathcal{D}_{t_e})$ were inserted before $\texttt{A}(\boldsymbol{x})$, it must have been in $\textit{dep}(\texttt{A}(\boldsymbol{x}) \mid \mathcal{D}_{t_b} - E),$ i.e., it does not violate P2E2.  
Consider the instantiated RDR in our running example $\delta^-_5$ ( E.g.~\ref{ex: reasoning}), although this is in $\textit{dep}(\schele{pLoc(4)} \mid \mathcal{D}_{4}),$ it does not violate P2E2 as this dependency on \schele{pLoc}(4) existed in the state $\mathcal{D}_1.$ 
We show that for an instantiated RDR to be in $\Delta^-(\textit{P2E2}, \texttt{A}(\boldsymbol{x}) )$ it must contain at least one attribute function that was inserted after $\texttt{A}(\boldsymbol{x}).$
\begin{theorem}
\label{thm: deletion_set}
    Let $\texttt{A}(\boldsymbol{x})$ be a cell with $\kappa(\texttt{A}(\boldsymbol{x})) = t_b$ and $\eta(\texttt{A}(\boldsymbol{x}))= t_e.$ Let $E$ be the set of erased cells between the states $\mathcal{D}_{t_b}$ and $\mathcal{D}_{t_e}.$ The following holds: (1)$\textit{dep}( \texttt{A}(\boldsymbol{x}) \mid \mathcal{D}_{t_b} - E\} \subseteq \textit{dep}( \texttt{A}(\boldsymbol{x} \mid \mathcal{D}_{t_b}\};$ (2) $\textit{dep}( \texttt{A}(\boldsymbol{x}) \mid \mathcal{D}_{t_e}) - \textit{dep}( \texttt{A}(\boldsymbol{x} \mid \mathcal{D}_{t_b} - E) = \{ \delta^-_i \mid \delta_i^- \in \textit{dep}(\texttt{A}(\boldsymbol{x}) \mid \mathcal{D}_{t_e}) \wedge \exists c \in \textit{Cells}(\delta^-_i) \text{ such that } \kappa(c) > t_b\}.$
        
\end{theorem}

Consider the timeline in Fig.~\ref{fig: timeline}, which shows the insertion of Bob's new post and other changes to the database. We want to guarantee P2E2 for the location of Bob's latest post ($\schele{pLoc}(\schele{4}))$. Not all the given dependencies hold for $ \schele{pLoc}(\schele{4})$. RDR R3 is not applicable in this case. Since the difference between the time of the post ($\schele{pTm}(\schele{4})$) and the time ($\schele{dTm}(\schele{1})$) at which the location of Bob's device ($\schele{dLoc}(\schele{1})$) was updated  
, $\schele{pLoc}(\schele{4})$ and $\schele{dLoc}(\schele{1})$ are not dependent. RDRs R2, R4, and R6 are applicable.

\noindent\textbf{Resolving Dependencies.}
What remains is to resolve the dependencies in $\Delta^-(\textit{P2E2}, \texttt{A}(\boldsymbol{x}) )$ to ensure that the P2E2 condition for $\texttt{A}(\boldsymbol{x})$ is satisfied. We want to identify a set $\mathcal{T}$ of cells in $\mathcal{D}_{t_e}$ such that when they are deleted(set to \textit{NULL}), P2E2 guarantee holds for $\texttt{A}(\boldsymbol{x}).$ We show that for each instantiated RDR that violates P2E2, we have to delete a cell from its head or tail.
\section{Supporting Demand-driven Erasure}
\label{sec: demand-driven}

\color{black}
In this section, we focus on demand-driven erasure. 
Given a set of RDRs and a cell $c_d$ for which P2E2 must hold, the first step is to instantiate the relevant RDRs (Sec.~\ref{sub: dep_inst_alg}). 
Our first approach reduces \textsc{Opt-P2E2} (Defn.~\ref{defn: P2E2}) to \textsc{ILP} (Sec.~\ref{sub: ILP}) which can be solved using readily available ILP solvers. Optimal answers obtained from solvers often have high overheads. We develop a hypergraph-based approach (Sec.~\ref{sub: dep_hyp}) that provides the optimal answer when RDRs are acyclic. Finally, we extend this to a heuristic approach (Sec.~\ref{sub: approx_app}) that guarantees P2E2, but not at the least cost.

\subsection{Instantiating RDRs}
\label{sub: dep_inst_alg}
Given a database state $\mathcal{D}_t$, and an instantiated attribute function $\texttt{A}(\boldsymbol{x}),$  we denote with $\Delta^-(\mathcal{D}_t, \texttt{A}(\boldsymbol{x}))$ the set of all instantiated relational dependency rules (RDRs) with $\texttt{A}(\boldsymbol{x})$ in the head or tail. All RDRs with $\texttt{A}(\boldsymbol{X})$ must be instantiated. Observe that to prevent \emph{direct} and \emph{indirect} inferences on $\texttt{A}(\boldsymbol{x})$, we need to consider instantiated RDRs in $\Delta^-(\mathcal{D}_t, \texttt{A}(\boldsymbol{x}))$ as well as all the instantiated attribute functions in $\bigcup\limits_{\delta^- \in \Delta^-(\mathcal{D}_t, \texttt{A}(\boldsymbol{x}))} \textit{Cells}(\delta^-)$  besides $\texttt{A}(\boldsymbol{x})$, and recursively so on. 

\begin{algorithm}
    \caption{Instantiating RDRs for P2E2}
    \label{alg: DepInstantiation}
    \begin{algorithmic}[1]

        \Procedure{DepInst}{$\mathcal{D}_t, \Delta^-, \texttt{A}(\boldsymbol{x})$}
            \State $\mathcal{Q} \gets \{ \texttt{A}(\boldsymbol{x})\}$ \Comment{Queue of cells}
            \State $\mathcal{V} \gets \emptyset, \mathcal{I} \gets \emptyset$\Comment{Lists of instantiated attributes \& RDRs}
            \While{$\mathcal{Q} \neq \emptyset$}
                \State $\textit{attf} \gets \mathcal{Q}.\textit{pop}$
                \If{$\textit{attf} \notin \mathcal{V}$}
                    \State $\mathcal{V} \gets  \mathcal{V} \cup \{\textit{attf}\}, Rules \gets \Delta^-(\textit{attf})$
                    \For{$rule \in Rules$}
                        \State $\delta^- \gets eval(rule), \mathcal{I} \gets \mathcal{I} \cup \{\delta^-\}$
                        \For{$\textit{tail} \in \textit{Tail}(\delta^-)$}
                            \State $\mathcal{Q}.\textit{push}(\textit{tail})$
                        \EndFor
                    \EndFor
                \EndIf
            \EndWhile
            \State \Return $\mathcal{V}, \mathcal{I}$ 
        \EndProcedure
    \end{algorithmic}
\end{algorithm}

Given $\texttt{A}(\boldsymbol{x})$ and a database $\mathcal{D}_t,$ Alg.~\ref{alg: DepInstantiation} generates the set $\Delta^-(\textit{P2E2}, \texttt{A}(\boldsymbol{x}) )$ of instantiated RDRs. The algorithm iteratively instantiates RDRs corresponding to direct or indirect inference of $\texttt{A}(\boldsymbol{x})$. 
The $\textit{eval()}$ function (in line~9) evaluates the condition of the rule, determines if the rule could have been used for inference in the state $\mathcal{D}_{\kappa(\texttt{A}(\boldsymbol{x}))^-}$ (Thm.~\ref{thm: deletion_set}) and returns the instantiated RDRs.

Alg.~\ref{alg: DepInstantiation} performs a breadth-first traversal over the attribute function dependency graph, starting from a target cell $\texttt{A}(\boldsymbol{x})$. Let $u$ be the number of unique attribute functions on which $\texttt{A}(\boldsymbol{x})$ is dependent and $r = |\Delta^-|$ the number of RDR templates. In the worst case, each of the $u$ attributes may match up to $r$ rules, each involving up to $a$ attribute functions. Thus, the total number of instantiated rules is $O(ur)$, and the number of enqueued attributes is $O(ura)$. Since each attribute is visited at most once and each rule is evaluated once per attribute, the overall time complexity is $O(ura)$, and the space complexity is $O(u + ur)$. In practice, the algorithm is efficient when the dependency graph is sparse and rule arity is small.


\subsection{ILP-Approach}
\label{sub: ILP}
We present a reduction from the \textsc{Opt-P2E2} problem to integer linear programming (ILP). To provide an intuition of the reduction, we introduce some concepts and notation. 
\begin{definition}[Induced Bipartite Graph]
   For a cell $\texttt{A}(\boldsymbol{x})$ in a state $\mathcal{D}_t,$ given the set $\Delta^-(\textit{P2E2}, \texttt{A}(\boldsymbol{x}) ) = \{ \delta^-_1, \ldots, \delta^-_n \}$ of instantiated RDRs, we define the induced bipartite graph
   $\mathcal{B}(\Delta^-(\mathcal{D}_t, \textit{P2E2}(\texttt{A}(\boldsymbol{x})))) = (V = V_L \cup V_R,\, E = E_H \cup E_T)$ where $V_L=  \{\delta^-_i | 1 \leq i \leq n \}$ and $V_R = \{ c_j \mid c_j \in \textit{Cells}(\delta^-_i) \}$ are the bipartition of the vertex set $V$. The set $E = E_H \cup E_T$ of edges contains, for every $\delta^-_i = c_{i_1} \dep  c_{i_2}, \ldots, c_{i_{n_i}}$, an edge $(\delta_i, c_{i_1}) \in E_H$ and for, $2 \leq j \leq n_i$, edges $(\delta_i, c_{i_j}) \in E_T.$
\end{definition}

\noindent \textbf{Reduction.} Consider, for a database state $\mathcal{D}_t$, a cell $\texttt{A}(\boldsymbol{x})$, denoted with $c_d$, for which P2E2 must hold, and for the set $\Delta^-(\mathcal{D}_t, \texttt{A}(\boldsymbol{x}))$ of instantiated RDRs, the induced bipartite graph $\mathcal{B}(\Delta^-(\mathcal{D}_t, \texttt{A}(\boldsymbol{x}))) = (V_L \cup V_R, E_H \cup E_T)$. We introduce the following variables: 
for $\delta_i \in V_L$ a binary variable $b_i$; for each $c_j \in V_R,$ a binary variable $a_j$; for each edge $(\delta_i, c_j) \in E_H,$ a binary variable $h_i^j$; for each edge $(\delta_i, c_j) \in E_T$ a binary variable $t_i^j.$ 
We specify the constraints in the following. 
\begin{enumerate}
   \item For P2E2 to hold for unit $c_d,$ it must be erased. So, $a_d = 1.$
   \item For each unit $c_j,$ where $1 \leq j \leq m,$ that needs to be erased, all instantiated RDRs where $c_j$ is in the head can be used to infer it. To prevent this, we need to address the instantiated RDR. This is stated, for all $i$, using the constraints $a_j = h^j_i.$
   
   \item For each instantiated RDR $\delta_i,$ if the unit in its head is hidden, then we set $b_i = h^i_j.$

   \item For an instantiated RDR $\delta_i,$ where $1 \leq i \leq n,$ to prevent the inference of the unit in its head, a unit from the tail has to be erased. So we introduce the constraint $\sum\limits_{j \in \{i_1, \ldots, i_{n_i}\}} t_i^j \geq b_i.$ 

   \item For each unit $c_j,$ for $1 \leq j \leq m,$ if the unit is erased, then all of the tail edges incident on it must indicate so which is ensured by, for all $1 \leq i \leq n,$ we have $a_j = t^j_i.$

   \item \label{opt_eqn}
   $ W = \texttt{min} \sum_{j=1}^{m} a_j$ minimizes the number of units erased. 
\end{enumerate}
Let the ILP system above be $\mathcal{O}.$ It has $O(nm)$ binary variables and $O(mn)$ constraints. For a cost model where, for $1 \leq j \leq m,$ the cost of erasing cell $c_j$ is $\textit{Cost}(j) \in \mathbb{N},$ we use 
$W = \texttt{min} \sum_{j=1}^{m} c_j \textit{Cost}(j)$ to minimize the total cost of P2E2-guarantee.
\color{blue}
\color{black}
\subsection{Dependence Hypergraph} 
\label{sub: dep_hyp}
Here, we present a new approach to solve \textsc{Opt-P2E2}. Given a set of instantiated RDRs $\Delta^-(\textit{P2E2}, \texttt{A}(\boldsymbol{x}) )$ we construct a \emph{dependence hypergraph} where the vertices are cells and a hyperedge connects, for an instantiated RDR $\delta^-$, its head $\textit{Head}(\delta^-)$ to its tail $\textit{Tail}(\delta^-)$.



\begin{definition}[Dependence Hypergraph]
    For a cell $\texttt{A}(\boldsymbol{x})$ in a database state $\mathcal{D}_t,$ given the set $\Delta^-(P2E2, \texttt{A}(\boldsymbol{x}))$ of instantiated RDRs, we define the \emph{dependence hypergraph} $\mathcal{H}(\Delta^-(\textit{P2E2}, \texttt{A}(\boldsymbol{x}) ))$ $= (V, E)$ where $V = \bigcup_{\delta^-_i \in \Delta^-(\mathcal{D}_t)} \textit{Cells}(\delta^-_i)$ is the set of vertices and $E = \{ (\textit{Head}(\delta^-_i), \textit{Tail}(\delta^-_i)) \mid \delta^-_i \in \Delta^-(\mathcal{D}_t)\}$ is the set of edges. 
\end{definition}
\begin{figure}[ht]
    \centering
    \includegraphics[width=3.5cm]{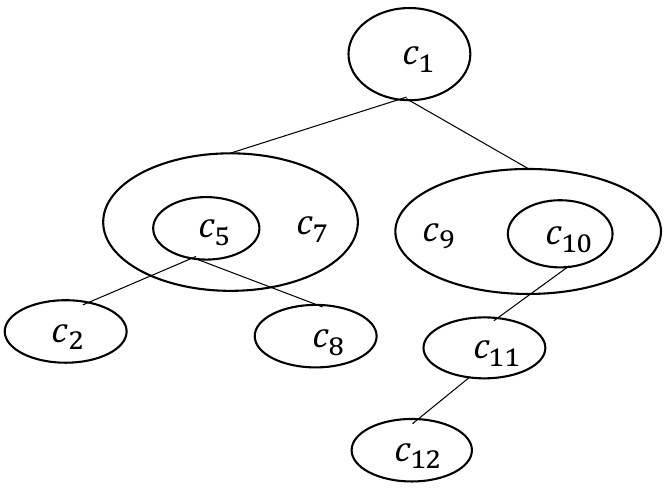}
    \Description{Example of a dependence hypergraph.}
    \caption{Example of a dependence hypergraph.}
    \label{fig:hyper_ex}
\end{figure}
For a dependence hypergraph $\mathcal{H}(\Delta^-(\textit{P2E2}, \texttt{A}(\boldsymbol{x}) )) = (V, E)$, a vertex $v \in V$ is called a \emph{root} if $v$ is not in the tail of any instantiated RDR, i.e., for all $\delta_i \in \Delta^-(\mathcal{D}_t)$ we have $v \notin \textit{Tail}(\delta_i).$ Similarly, a vertex $v \in V$ is called a \emph{leaf} if $v$ is not the head of any instantiated RDR, i.e., for all $\delta_i \in \Delta^-(\mathcal{D}_t)$ we have $v \notin \textit{Head}(\delta_i).$
Fig.~\ref{fig:hyper_ex} shows the  dependence hypergraph for the set $\{\delta^-_1 : c_1 \dep c_5, c_7;~ \delta^-_2 : c_5 \dep c_2;~ \delta^-_3 : c_5 \dep c_8;~ \delta^-_4 : c_1 \dep c_9,c_{10};~ \delta^-_5: c_{10} \indep c_{11};~ \delta^-_6: c_{11} \indep c_{12} \}.$ 
The vertex $c_1$ is a root and $c_9$ and $c_8$ are leafs. 
Next, we define paths and complete paths in a dependence hypergraph to characterize the P2E2-guarantee.

\begin{definition}
    For a dependence hypergraph $\mathcal{H}(\Delta^-(P2E2, \texttt{A}(\boldsymbol{x})))$, and a sequence $P: v_1, v_2, \ldots, v_{n}$ of vertices, we define the following.
\begin{itemize}
    \item We say that the sequence $P$ is a \emph{path} if the following hold:
    \begin{enumerate}
        \item There exists $\delta_i$ s.t. $v_1 \in \textit{Head}(\delta_i).$
        \item For $1 \leq i < n,$ there exists $\delta_i$ s.t. \\
            $v_i \in \textit{Head}(\delta_i)$ and $v_{i+1} \in \textit{Tail}(\delta_i)$
    \end{enumerate} 
    
    \item We say that a sequence $P': v_b, \ldots, v_e$ where $1 \leq b \leq e \leq n,$ is a \emph{subpath} of $P$ if $P$ and $P'$ are both paths. A path is, trivially, a subpath of itself.
    
    \item We say that the sequence $P$ is a \emph{complete (sub-)path} if $P$ is a (sub-)path such that $v_n$ is a leaf. 
\end{itemize}
\end{definition}

 In Fig.~\ref{fig:hyper_ex}, the vertex $c_1$ is a root, vertices $c_7 \text{ and } c_8$ are leaves; the sequence  $P_1: c_1, c_{10}, c_{11}$ is a path; the sequence $P_2: c_1, c_5, c_2$ is a complete path; $P_3: c_5, c_2$ is a subpath of $P_2.$

We can now use the definition of path above to characterize when P2E2 holds. In particular, if there exists a path in $\mathcal{H}(\Delta^-(P2E2, \texttt{A}(\boldsymbol{x})))$ such that for all vertices on the path,
there is a subpath. We show that this is both a necessary and a sufficient condition for P2E2. Note that simply defining a cover~\cite{khot2008vertex} would have been sufficient but not necessary for P2E2 to hold.  
In the dependence hypergraph in Figure~\ref{fig:hyper_ex}, it suffices when the units $c_1, c_7, \text{ and } c_9$ are set to $\textit{NULL}$ for P2E2 to hold (but they do not cover the graph).

\begin{algorithm}
    \caption{Optimizing the Dependence Hypergraph}
    \label{alg: Hypergraph}
    \begin{algorithmic}[1]
        \Procedure{OptPath}{$\mathcal{V}, \mathcal{I}$} \Comment{Output of Algorithm 1}
            \State $\mathcal{Q} \gets \{\textit{leafs} (\mathcal{V})\},$ $\mathcal{S} \gets \emptyset$ \Comment{Queue of cells and set of seen cells}
            \While{$\mathcal{Q} \neq \emptyset$}
                \State $\textit{attf} \gets \mathcal{Q}.\textit{pop}$
                \If{$\textit{attf} \notin \mathcal{S}$}
                    \State $\mathcal{S} \gets \mathcal{S} \cup \{\textit{attf}\}$
                    \State $\textit{Cost}(\textit{attf}) \gets 1$ \Comment{Initialize node cost}
                    \label{alg: Hypergraph: base cost}
                    \State $Rules \gets \mathcal{I}(\textit{attf})$ \Comment{All RDRs that contain \textit{attf}}
                    \For{$\delta^- \in Rules$}
                         \If{$\textit{attf} \in \textit{Head}(\delta^-)$}
                            \State $\textit{Sum}(\textit{Cost}(\textit{attf}), min ~ \textit{Cost}(\textit{Tail}(\delta^-)))$ \Comment{\\Add cheapest node of tail to node's cost}
                            \label{alg: Hypergraph: sum cost}
                         \Else{}
                            \State $\mathcal{Q}.\textit{push}(\textit{Head}(\delta^-))$ \Comment{Add RDR head to queue}
                         \EndIf
                    \EndFor
                \EndIf
            \EndWhile
            \State $\mathcal{T} \gets \{\texttt{A}(\boldsymbol{x})\}, \mathcal{Q} \gets \{\texttt{A}(\boldsymbol{x})\},$ $\mathcal{S} \gets \emptyset$ \Comment{Set of cells to delete}
            \While{$\mathcal{Q} \neq \emptyset$}
                \State $\textit{attf} \gets \mathcal{Q}.\textit{pop}$
                \If{$\textit{attf} \notin \mathcal{S}$}
                    \State $\mathcal{S} \gets \mathcal{S} \cup \{\textit{attf}\}, Rules \gets  \mathcal{I}(\textit{attf})$
                    \For{$\delta^- \in Rules$}
                        \If{$\textit{attf} \in \textit{Head}(\delta^-)$}
                            \State $\textit{child} \gets arg\,min ~ \textit{Cost}(\textit{Tail}(\delta^-))$
                            \label{alg: Hypergraph: min cost}
                            \State $\mathcal{T} \gets \mathcal{T} \cup \textit{child}$
                            \State $\mathcal{Q}.\textit{push}(\textit{child})$ \Comment{Delete cheapest node in tail and continue traversal}
                        \EndIf
                    \EndFor
                \EndIf
            \EndWhile
            \State \Return $\mathcal{T}$
        \EndProcedure
    \end{algorithmic}
\end{algorithm}
\smallskip \noindent \textbf{Optimization.} Given a set $\Delta^-$ of RDRs, a database state $\mathcal{D}_t$, an instantiated attribute function $\texttt{A}(\boldsymbol{x})$, Algorithm~\ref{alg: DepInstantiation} can be easily adapted to construct a dependence hypergraph. 
Given a dependence hypergraph,  Algorithm~\ref{alg: Hypergraph} produces an optimal solution. We assume that the graph is cycle-free. This assumption is necessary only for proving optimality\footnote{If cycles are present, our approach produces a correct solution but not necessarily an optimal one. To ensure that cycles are not present in the hypergraph, for all pairwise instantiated dependencies $\delta^-_1$ and $\delta^-_2$, if $\textit{Tail}(\delta^-_1) \cap \textit{Tail}(\delta^-_2) \neq \emptyset$, discard the instantiated RDR with the larger number of attribute functions in its tail. The solution produced by Algorithm~\ref{alg: Hypergraph} is optimal with respect to the thus obtained set of RDRs.}.

In practice, we implement the following procedure.
The cost of every node is set to~1, as a base cost to erase a single attribute function (Line~\ref{alg: Hypergraph: base cost}).
The \emph{leafs} cannot incur a higher cost, as they do not cause additional erasures.
Next, we traverse the tree upwards and compute the cost of every \emph{inner node} as the sum of the nodes' cost and the minimal cost of every attached hyper edge (Line~\ref{alg: Hypergraph: sum cost}).
When reaching the root node, we construct the complete path by traversing the tree to the bottom and always choosing the node with minimal cost (Line~\ref{alg: Hypergraph: min cost}).
Thus, the algorithm guarantees P2E2 at minimal cost but has to traverse the tree twice.

\begin{theorem}
\label{thm:correctnessHyG}
    When Algorithm~\ref{alg: Hypergraph} terminates, the set $\mathcal{T}$ contains cells which, when set to $\textit{NULL},$ guarantees P2E2 for the input $\texttt{A}(\boldsymbol{x}).$ Moreover, $\sum_{\texttt{A}_i(\boldsymbol{x}_i) \in \mathcal{T}} \textit{Cost}(\texttt{A}_i(\boldsymbol{x}_i))$ is minimized.
\end{theorem}

The algorithm consists of two phases: a bottom-up cost propagation to assign minimal deletion costs, followed by a top-down traversal to extract the optimal deletion set. Let $n = |\mathcal{I}|$ be the number of instantiated RDRs, $a$ the maximum rule arity, and $C_{\max}$ the maximum cell cost. The bottom-up phase runs in $O(n \cdot a \cdot \log C_{\max})$ time, and the top-down phase adds $O(C_{\max} \cdot a \cdot \log C_{\max})$ for path extraction. Hence, the total time complexity is $O\left((n + C_{\max}) \cdot a \cdot \log C_{\max}\right)$, with space complexity $O(n \cdot a)$. In practice, the algorithm is efficient due to typically low arity and shallow dependency paths.

\color{black}
\subsection{Approximate Algorithm}
\label{sub: approx_app}
In this section, we present an approximation variant of Algorithm~\ref{alg: Hypergraph}, which, given a cell, determines a (not necessarily the smallest) set of dependent cells to delete for P2E2 to hold. 

We adapt the algorithm to determine the minimum cost of guaranteeing P2E2 (Alg.~\ref{alg: Hypergraph}) such that instead of constructing the entire dependence hypergraph and then traversing it bottom-up, a partial top-down construction of the dependence hypergraph is sufficient: whenever there is an instantiated RDR that has more than two attribute functions, we only instantiate the next \emph{level} in the tree and choose the one that has the lowest cost to erase. The other attribute functions are not instantiated fully, thereby saving time.

The algorithm greedily constructs a partial dependence hypergraph top-down and avoids the bottom up traversal as in Alg.~\ref{alg: Hypergraph}. Since each of the $n$ RDRs (each with maximum arity $a$) is instantiated at most once for each of its $a$ cells, the total number of instantiated cells is $O(an)$. As the hypergraph is traversed only once, the time complexity is $O(an)$ and the space complexity is $O(an)$.

Observe that greedily selecting the attribute function with the lowest deletion cost can lead to suboptimal outcomes—for example, forcing the deletion of a high-cost function later. 
Let $\mathcal{T}$ denote the deletion set from the greedy algorithm and $\mathcal{T}^*$ the optimal set. In general, the cost ratio $\frac{\textit{Cost}(\mathcal{T})}{\textit{Cost}(\mathcal{T}^*)}$ can be unbounded. However, when minimizing the cardinality of the deletion set, if the number of cells per instantiated RDR is bounded by arity $a = \max_{\delta^- \in \Delta^-(\textit{P2E2}, \texttt{A}(\boldsymbol{x}) )} |\textit{Cells}(\delta^-)| \geq 2,$ if a cell is in at most $d$ instantiated RDRs,
i.e., $d = \max_{c' \in \cup_{\delta^- \in \Delta^-(\textit{P2E2}, \texttt{A}(\boldsymbol{x})) }\textit{Cells}(\delta^-)} |\{ \delta^-| c' \in \textit{Cells}(\delta^-)\}|,$ and the number of cells in the acyclic graph is $n, $ then we can bound the approximation ratio as follows.

\begin{theorem}
\label{thm:uniform_apx}
  Given an acyclic dependence hypergraph \\ $\mathcal{H}(\Delta^-(\textit{P2E2}, \texttt{A}(\boldsymbol{x}) ))$$= (V, E),$ with $|V|=n$ and $|\Delta^-(\textit{P2E2}, \texttt{A}(\boldsymbol{x}) )|=r,$ let $\mathcal{T}^*$ be the minimal set of cells that need to be deleted to guarantee P2E2 for $\texttt{A}(\boldsymbol{x})$ and $\mathcal{T}$ be the set of cells to be deleted to guarantee P2E2 with the approximation algorithm (in Sec.~\ref{sub: approx_app}).  The following holds: $\frac{|\mathcal{T}|}{|\mathcal{T}^*|} = \min\{ \frac{d}{r} \cdot \left( 1 + \log_2\left( \frac{a \cdot r}{d} \right) \right), \frac{ad}{n-1}(1 + \log_2(an))\}
.$
\end{theorem}


\subsection{Batching Erasures}
\label{sub: batching}
The approaches discussed until now focus on the erasure of one cell. Since regulatory data erasure requirements allow for a reasonable delay between the time at which the data is requested to be erased and the actual erasure of the data (referred to as grace period and denoted with $\Gamma$), it is possible to batch data erasures. The grace period can be used to batch multiple data erasure requests and instead of constructing and solving an individual optimization model for each cell, we attempt to construct models that allow for multiple cells to be erased such that the P2E2 holds for each of them.

Intuitively, we instantiate RDRs for the cells to be erased, which maximizes the possibility that the corresponding dependence hypergraphs have shared vertices. In practice, over a $\Gamma$ period of time, we collect all cells that have to be erased such that P2E2 holds for them. Let this set of cells be $S$. We instantiate RDRs for each cell $s\in S$ at a time. Whenever an instantiated RDR corresponding to $s$ contains a dependent cell $s'$ also in $S$, we mark it to be set to $\textit{NULL}$ and only instantiate the RDRs for~$s'$. This not only reduces the number of RDRs instantiated and, thus, the number of leafs in the tree, but also the time taken to traverse the tree. Moreover, fewer models (ILP or hypergraphs) need to constructed and optimized.  

\section{Retention-driven Erasure}
\label{sec: retention}
So far, we have considered demand-driven erasures (a user wants to erase a cell $c$ before its expiration time $\eta(c))$ and batching such erasures. Now we turn to retention-driven erasure (where cell $c$ is erased at its preset $\eta(c)).$ 
For such erasures, we investigate how to minimize the overheads of P2E2 on derived data. For base data, we adopt the batching approach discussed in Sec.~\ref{sub: batching}.

Guaranteeing P2E2 for cells, often requires additional and potentially undesirable update and reconstruction of derived data. For example, suppose, for a derived cell $c$, with the parameter $\textit{freq}(c) = 1 \textit{hr},$ depends on cells $c_1, c_2, \text{ and } c_3.$ It is reconstructed at 1pm, 2pm, 3pm, and so on. The cells $c_1, c_2, \text{ and } c_3$ expire at 1:30pm, 3:00pm, and 4:30pm, respectively. To guarantee P2E2 for the dependent cells, cell $c$ needs to be reconstructed at 1pm, 1:30pm, 2:30pm, 3pm, 4pm, and at 4:30pm thus incurring additional overheads.

Retention-driven erasures offer an opportunity to reduce additional reconstructions due to P2E2
by exploiting the already known expiration times.
We present an algorithm that, given a derived cell~$c$ and its dependencies $c_1, \ldots, c_n$ with corresponding erasure time intervals\footnote{Erasure time interval refers to the time interval in which a cell has to be erased. Usually $\eta^b_i + \Gamma = \eta^e_i$. However, here we allow for cells to have different time intervals in which they must be erased.} $(\eta^b_1, \eta^e_1), (\eta^b_2, \eta^e_2), \ldots, (\eta^b_n, \eta^e_n)$, determines an erasure schedule $Sch(c)$ that takes into account when derived data has to be erased while maintaining the invariant that $c$ is reconstructed at least once every $\textit{freq}(c)$. Intuitively, we progressively build the reconstruction schedule $Sch(c)$ by determining the maximum overlap between the retention periods of the dependent data
to minimize the number of extra reconstructions due to P2E2. 

\begin{algorithm}
    \caption{Reconstruction Scheduler}
    \label{alg: scheduler}
    \begin{algorithmic}[1]
        \Procedure{createSchedule}{$c, \textit{freq}(c), \textit{depSet}$}
            \State $\kappa(c) \gets time.now()$
            \State $\textit{Sch}[0] \gets \textsc{maxOverlap}(\textit{depSet}, \kappa(c) + \textit{freq}(c))$
            \State $\textit{depSet} \gets \textit{depSet} \setminus \{c_i \mid \textit{Sch}[0] \in (\eta^b_i, \eta^e_i) \}$
            \State $i \gets 1$
        \While{$\textit{depSet} \neq \emptyset$}
            \State $\textit{Sch}[i] \gets \textsc{maxOverlap}(\textit{depSet}, \textit{Sch}[i-1] + \textit{freq}(c))$ 
            \State $\textit{depSet} \gets \textit{depSet} \setminus \{c_i \mid Sch[0] \in (\eta^b_i, \eta^e_i) \}$
            \State $i \gets i + 1$
        \EndWhile
        \EndProcedure
      \Procedure{updateSchedule}{$c, \textit{freq}(c), \textit{Sch}, \textit{depSet}$}
            \For{$(c_i, \eta^b_i, \eta^e_i) \in \textit{depSet}$}
                \If{$\eta^b_i > \textit{Sch}[0] \wedge \eta^e_i < \textit{Sch}[1]$}
                \State $\textsc{createSchedule}(c, \textit{freq}(c), \textit{depSet})$
                \EndIf
            \EndFor
        \EndProcedure
    \end{algorithmic}
\end{algorithm}

Our algorithm (Algorithm~\ref{alg: scheduler}) is in two parts: Step~1 (Lines 1-11) creates the reconstruction schedule $\textit{Sch}(c)$ of the cell $c$, and Step~2 (Lines 12-18) updates the schedule when required to ensure that newly inserted dependencies are accounted for. At any given time, $\textit{depSet}$ for a derived cell $c$ denotes the set $\{(c_i, \eta^b_i, \eta^e_i) \mid 1 \leq i \leq n \}$ of all its dependencies, their insertion time, and their erasure time, respectively. The \textsc{maxOverlap} function is a standard algorithm to find the maximum overlap given a set of time intervals.

\noindent Step~1: The first step finds a reconstruction time $\rho$ that maximizes for $1 \leq i \leq n$ the overlap between the erasure time intervals $(\eta^b_i, \eta^e_i)$  of the dependent cells, and the time interval  $(\kappa(c), \kappa(c) + \textit{freq}(c))$ in which $c$ must be reconstructed at least once. Therefore, P2E2 holds for any cell $c_i$ where $\rho \in (\eta^e_i, \eta^e_i)$. The algorithm iteratively finds the maximum overlap and creates a list $Sch(c): \rho_1, \ldots, \rho_m$ of reconstruction times for $c$ such that P2E2 holds for its dependencies in the database at the time of the construction of the schedule.

\noindent Step 2: The update procedure is called when a derived cell $c$ is reconstructed. It checks whether there exists a dependent cell $c_i$ that has to be erased after the current reconstruction but before the next scheduled reconstruction. If it does, then a new reconstruction schedule is created using the first step described above. Observe that if the erasure time of any dependent cell $c_i$ is updated, this step (Step~2) ensures that P2E2 holds. 

Let $n$ be the number of dependent base cells for a derived cell $c$. 
Since the algorithm utilizes a greedy interval scheduling problem, the time complexity is $O(n \log n)$ due to sorting, and the space complexity is $O(n)$ to store intervals and the resulting schedule.

\section{Evaluation}
\label{sec: eval}
We evaluate our approaches for guaranteeing P2E2.
We compare the ILP approach (Sec.~\ref{sub: ILP}) with the graph-based algorithm, \textsc{HGr}~(Sec.~\ref{sub: dep_hyp}) and its approximate version, \textsc{Apx}~(Sec.~\ref{sub: approx_app}).
We analyze the efficiency and effectiveness of all the algorithms when applied to individual demand-driven erasures as well as with a set of erasures using our batching method, \textsc{Batch}~(Sec.~\ref{sub: batching}).
Additionally, we investigate our approach for retention-driven erasures, \textsc{Scheduler}~ (Alg.~\ref{alg: scheduler}).

\subsection{Experimental Setup}
All experiments were run on an Ubuntu-based (20.04 LTS) server (Intel Xeon E5-2650; RAM: 256~GB).
All algorithms are single-threaded, running on Java~11 and the datasets are stored in a PostgreSQL (v12.20) database. The ILP approach uses the Gurobi (v11.03) solver.

%

\noindent
\textbf{Datasets.} 
We use the following five datasets in our evaluations (the first four columns of Table~\ref{tab:eval:instantiated} summarizes them and shows the number of RDRs and where/how they were derived.) 
\noindent
(1) \emph{Twitter}~\cite{twitter100m_tweets}. This dataset contains a subset of tweets posted on $\mathbb{X}$ (formerly Twitter) that represents a real-world instance of our running example.
The RDRs were designed manually and express dependencies between individuals and their content.\\
(2) \emph{Tax}~\cite{bohannon2007conditional}. This dataset is a synthetic dataset created using the real-world distribution of values of American tax records. 
We discovered all present DCs using~\cite{pena2019discovery}.
The top-10 DCs that are not entirely comprised of equality predicates are transformed into RDRs. The RDRs include the conditional functional dependencies used in the original publication for data cleaning.
\\
(3) \emph{SmartBench}~\cite{gupta2020smartbench}. This dataset is based on real data collected from sensors deployed throughout the campus at the University of California, Irvine.
RDRs capture dependencies between multiple physical sensors which are used to compute derived metrics, e.g., occupancy from Wi-Fi AP locations.
\\
(4) \emph{HotCRP}~\cite{HotCRP}. This is a dataset containing a sample of real-world conference data. It stores authors, papers, conferences, and their relationships.
The RDRs are generated by the method of~\cite{GDPRizer}. 
\\
(5) \emph{TPC-H}~\cite{TPC-H}. This well-known benchmark dataset stores transactions of commercial actors: customers and their orders, as well as suppliers that fulfill those.
The RDRs capture the links between the tables, i.e., foreign keys.
In TPC-H, both customers and suppliers may delete their data.We create two separate scenarios and combine them proportionally to the number of customers and suppliers.
\\
The set of RDRs for each dataset is cycle free.
Some RDRs in the Twitter and SmartBench dataset join on non-key columns.
To speed up the instantiations, we index those columns separately. 

\noindent\textbf{Metrics.} We measure the (i)~total number of deletions, (ii)~time taken, and (iii)~space overheads to guarantee P2E2. For demand-driven erasure, we also measure the number of reconstructions. 

\noindent
\textbf{Workload.} Given the lack of suitable deletion benchmarks \cite{Lethe}, we evaluate demand- and retention-driven erasures separately, as well as varying combinations of each.

\subsection{Experiments}
\label{sub: experiments}

 \noindent \textbf{Experiment 1.}
\begin{figure*}[t]
    \centering
    \scriptsize
    \begin{subfigure}{.90\linewidth}\setlength{\tabcolsep}{4pt}
        \centering
        \begin{tabular}{@{}lrrrrrrrrr|rrr@{}}
            \toprule
            Dataset & \# Cells & \# RDRs & \# base, & \multicolumn{3}{c}{Instantiated cells} & \multicolumn{3}{c|}{Deleted cells} & \multicolumn{3}{c}{Deleted cells for Baselines} \\ 
            & & (Source) &\# derived & ILP & \textsc{HGr} & \textsc{Apx} & ILP & \textsc{HGr} & \textsc{Apx} & \textsc{Inst} & \textsc{OpR} & \textsc{MinSet} \\ \midrule
            $\mathbb{X}$ (Twitter)\footnotemark & \num{21926096} & \num{11}$^*$ & \num{7}, \num{11}  & \num{0.29} & \num{0.29} & \num{0.29} & \num{0.29} & \num{0.29} & \num{0.29} & \num{837.21} & \num{517.8}  & \num{321.1} \\
            Tax~\cite{bohannon2007conditional} & \num{16000004} & \num{10} (\cite{pena2019discovery}) & \num{15}, \num{4}  & \num{6.93} & \num{6.93} & \num{5.14} & \num{4.07} & \num{4.07} & \num{4.2} & \num{6.93} & \num{6.6} & \num{4.07} \\
            SmartBench~\cite{gupta2020smartbench} & \num{94424} & \num{5}$^*$ & \num{5}, \num{1} & \num{5.43} & \num{5.43} & \num{5.43} & \num{5.43} & \num{5.43} & \num{5.43} & \num{1021.8} & \num{1021.8} & \num{1021.8} \\
            HotCRP~\cite{HotCRP} & \num{122697} & \num{3}(\cite{GDPRizer}) & \num{6}, \num{3} & \num{134.4} & \num{134.4} & \num{4.81} & \num{3.78} & \num{3.78} & \num{3.78} & \num{815.0} & \num{721.7} & \num{74.5} \\
            TPC-H~\cite{TPC-H} & \num{5825639} & \num{6} (\cite{gilad2020multiple}) & \num{8}, \num{4} & \num{17.63} & \num{17.63} & \num{17.63} & \num{17.63} & \num{17.63} & \num{17.63} & \num{722.2} & \num{722.2} & \num{722.2} \\
            \bottomrule
        \end{tabular}
        \caption{Summary of datasets and average number of instantiated \& deleted cells. \textsc{Inst} = all instantiated cells, \textsc{OpR} = one cell per rule, \textsc{MinSet} = minimal set, {$^*$}manually designed.}
        \label{tab:eval:instantiated}
    \end{subfigure}
     \centering
    \begin{subfigure}{\linewidth}
        \centering\includegraphics[width=.9\linewidth]{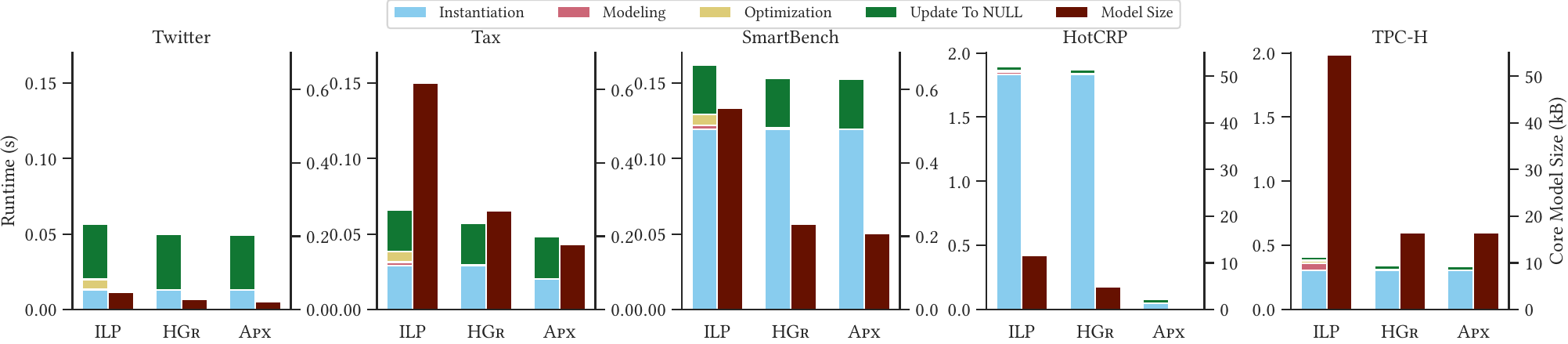}
        \caption{Average runtime and space overhead}
        \label{fig:eval:demanddriven}
    \end{subfigure}
    \Description{Evaluation of demand-driven erasures.}
    \caption{Evaluation of demand-driven erasures}
\end{figure*}
To test the cost of demand-driven erasures, we erase \num{100} random cells for each base data attribute in the RDRs.
We depict the average runtime and model size for a single erasure in Fig.~\ref{fig:eval:demanddriven}.
The total runtime is divided into the following four steps: 
(i)~RDR instantiation (Algorithm~\ref{alg: DepInstantiation});
(ii)~model construction (graph construction in \textsc{HGr} and \textsc{Apx} and defining the ILP instance);
(iii)~model optimization (traversing the graph and keeping track of the minimal cost erasure for \textsc{HGr} and solving the ILP; there is no optimization phase in \textsc{Apx}, as it greedily chooses the next edge to process);
(iv)~update to \textit{NULL} (which modifies the database to guarantee P2E2.)
We measure the average number of instantiated and deleted cells beyond the initial erased cell and compare our three P2E2 algorithms against three baseline implementations inspired by cascading deletions~\cite{UllmanBook, gilad2020multiple}. These baselines guarantee P2E2 by identifying and deleting dependent cells but, unlike our approaches, they cannot account for the state of the database at the time of insertion of the data being deleted. Each baseline scans the entire dataset but varies in how it selects which cells to delete: \textsc{Inst} deletes all dependent (i.e., instantiated) cells, mimicking traditional cascading deletes;
 \textsc{OpR} improves this and deletes exactly one cell per instantiated RDR; finally, \textsc{MinSet} computes a minimal set of cells such that at least one cell is deleted for each RDR (see Table~\ref{tab:eval:instantiated}). 
 
 \textit{Comparison with baselines: }As expected, all three baseline methods delete significantly more data than our P2E2 algorithms on every dataset except Tax. This is because, to guarantee P2E2 for a cell, we only need to consider data that was inserted after the cell and that is dependent on it. In high-volume datasets like Twitter and SmartBench, where a user generates many entries, the overhead of baselines is especially pronounced. In contrast, the Tax dataset does not incur additional deletions under \textsc{MinSet} because all user data is inserted simultaneously.
\footnotetext{\url{https://huggingface.co/datasets/enryu43/twitter100m_tweets}}

\textit{Analysis of our P2E2 mechanisms: }We observe a stark difference between Twitter, Tax, SmartBench, and HotCRP and TPC-H.
We highlight the similarity within these groups by using the same axis scaling.
RDRs for the HotCRP and TPC-H datasets were created using data dependencies from schema constraints (or IND discovery).
DBMSs already include a mechanism to delete data linked by foreign keys.
Thus, there is no overhead to guarantee P2E2.

Interestingly, neither the number of rules, nor the dataset size determine the deletion complexity.
P2E2 is more sensitive to the amount of related data, i.e., the number of instantiated cells.
Consequently, the instantiation time takes the most amount of time for the SmartBench, HotCRP, and TPC-H dataset. 
 In contrast, the instantiation time of the Tax dataset is lower although the number of instantiated cells is higher than in the SmartBench dataset.
The RDRs in the Tax dataset specify connections only within one row of the data.
For each deleted cell, we repeatedly query the same row, which is faster than scanning larger parts of the data as observed in SmartBench.
In the Twitter dataset, the final update step dominates the cost, as the number of instantiated cells is low.

We observe that the ILP approach has the highest overheads (runtime and memory) for all three datasets.
However, it is optimal in that it guarantees P2E2 using a minimum set of additionally deleted cells.
Likewise, \textsc{HGr} always produces an optimal result, but consumes significantly less memory.
Both approaches need to instantiate all available RDRs for all applicable cells, so their instantiation time is identical.
However, the model construction and optimization overhead for \textsc{HGr} is negligible compared to the ILP approach.
\textsc{Apx} instantiates fewer cells, so it outperforms the other approaches for all datasets.
This behavior is especially noticeable for HotCRP.
The optimal models have to instantiate a long chain of RDRs that turn out to be irrelevant to identify the cheapest deletion set.
Due to its greedy nature, \textsc{Apx} avoids instantiating all those RDRs and significantly outperforms the rest of the algorithms.
Since it does not keep track of an erasure cost, it also consumes less memory.
However, it cannot guarantee optimality for its result set, as exemplified in the Tax dataset (see Table~\ref{tab:eval:instantiated}).



%
\begin{figure}[b]
    \centering
\includegraphics[width=.8\linewidth]{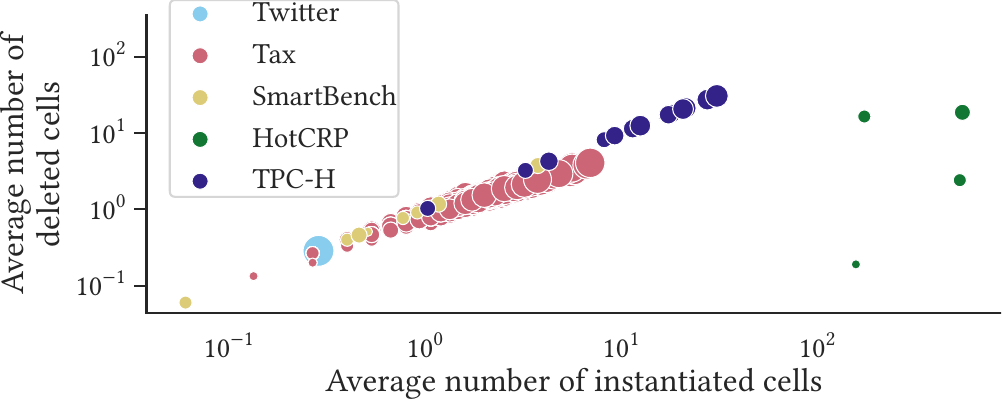}
     \Description{Impact of dependencies (log-axis) on P2E2 overhead; the size of data points signals number of RDRs}
    \caption{Impact of dependencies (log-axis) on P2E2 overhead; the size of data points signals number of RDRs}
    \label{fig:eval:dependence}
\end{figure}
\begin{figure*}[t]
    \centering
    \includegraphics[width=.9\linewidth]{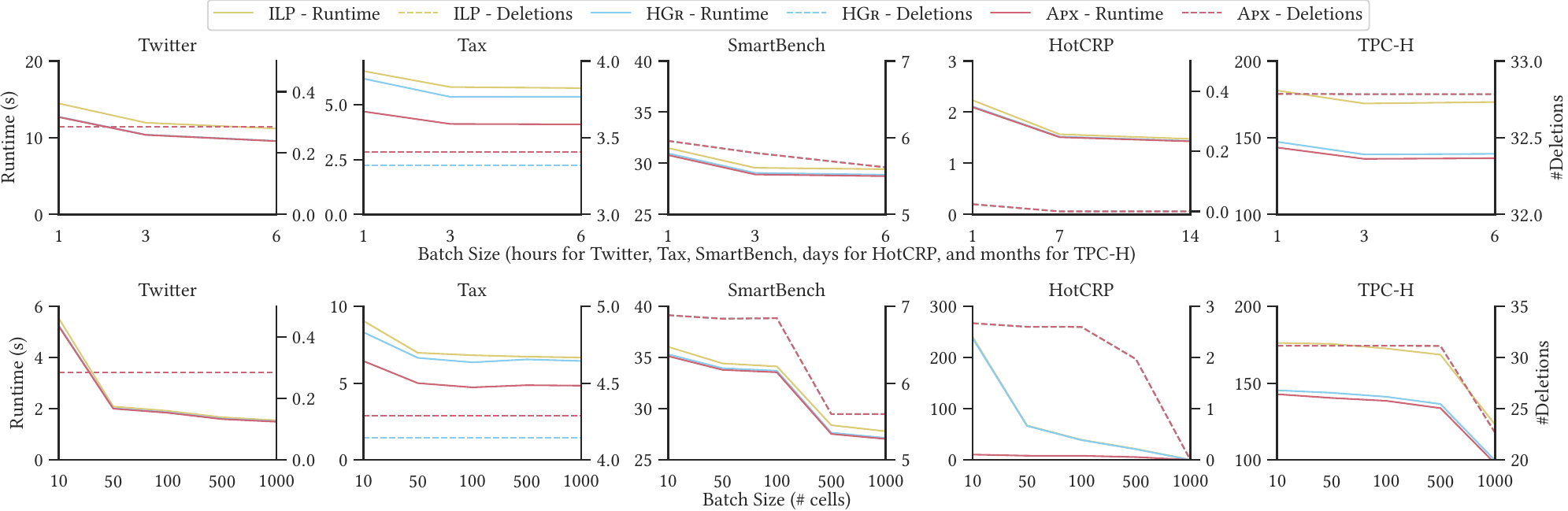}
    \Description{Evaluation of the batching.}
    \caption{Evaluation of the batching}
    \label{fig:eval:batching}
\end{figure*}

\noindent \textbf{Experiment 2.}To investigate the influence of the degree of dependence (determined by the count of the number of cells that are part of instantiated RDRs) in the data on the number of deleted cells, we conducted the following experiment.
For each dataset, we randomly sample 100 erasures.
Each sampled erasure is processed by \textsc{HGr} given every subset of RDRs.
In Fig.~\ref{fig:eval:dependence}, we plot the average number of instantiated cells compared to the average number of deleted cells.
The size of the data points signals the number of RDRs present in the processed subset.
Intuitively, the more dependent cells a cell has, the more needs to be deleted.
The number of rules does not have an immediate effect on the number of deleted cells, as larger and smaller points are mixed along the general trend line.
However, the HotCRP dataset is an outlier.
Due to the aforementioned long chain of instantiated RDRs, the number of instantiated cells is high, while the necessary deletions remain low.

\noindent \textbf{Experiment 3.}
We evaluate the impact of \textsc{Batch}, which exploits the grace period $\Gamma$ to batch as many erasures as possible.  
First, we batch erasures based on a time interval. The number of erasures in such a batching strategy is workload-dependent.
We further create batches based on fixed number of erasures to study the impact of batching across datasets in a workload-independent fashion.

\textit{Batching based on time:}
To unify the process for the Twitter and SmartBench datasets, we randomly sample \num{1000} erasures from a twelve-hour window in our data.
For the Tax dataset, we assume \num{100} record updates per hour.
After sampling the erasures, we use three different grace periods: one, three, and six hours.
The HotCRP and TPC-H dataset operate on a different timescale.
Therefore, we sample one month for the HotCRP dataset and one year for the TPC-H dataset.
They use the batch sizes of one, seven, fourteen days, and one, three, and six months, respectively.
The cumulated runtime and number of cells deleted for each batch size is depicted in Fig.~\ref{fig:eval:batching}.
In general, larger batch sizes require fewer cells to hide and are processed faster.
Initially, batching provides a larger benefit, as the impact of finding the first cells that are already instantiated is larger.
In the Twitter and Tax dataset, the number of additionally deleted cells stays constant.
In contrast, it scales similar to the runtime improvement in the HotCRP and TPC-H dataset.
While in the SmartBench dataset, the number of additionally deleted cells scales linearly, the runtime exhibits a steeper slope between the batch sizes of one and three hours.

\textit{Batching based on cardinality:}
We randomly sampled \num{1000} erasures from the entire dataset, and grouped them in batches of size \num{10}, \num{50}, \num{100}, \num{500}, and \num{1000}.
We present the runtime and number of deleted cells for the entire set of erasures in Fig.~\ref{fig:eval:batching}.
We observe a similar pattern as in the time-based method, larger batches perform better.
In the Twitter, Tax, and HotCRP datasets, the scaling trend is similar to the previous experiment.
For SmartBench, the biggest improvement occurs when enlarging the batch size from 100 to 500 cells.
As the erasures are sampled uniformly from the entire dataset, they are less likely to overlap in the instantiated cells and improve runtime.
Enlarging the batch size increases this likelihood.
Similarly, in the larger TPC-H dataset, this improvement only happens for the largest batch size.
In the HotCRP dataset, \textsc{HGr} and \textsc{ILP} are as performant as \textsc{Apx} because there is no need to instantiate the long chain of RDRs, if parts of it are already in the batch. 

\noindent \textbf{Experiment 4.}
\begin{figure}[b]
    \centering
    \includegraphics[width=.8\linewidth]{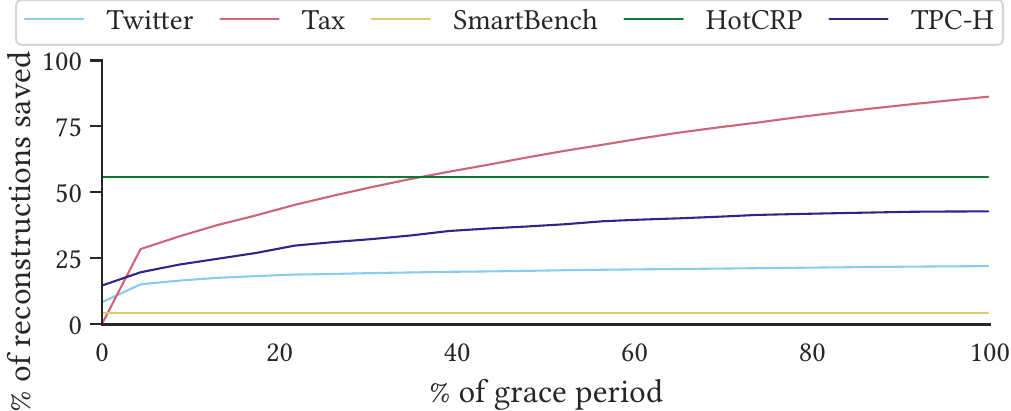}
    \Description{Number of saved reconstructions vs. grace period}
    \caption{Number of saved reconstructions vs.\ grace period}
    \label{fig:eval:scheduling}
\end{figure}
Here, we evaluate the effectiveness of \textsc{Scheduler} in reducing extra reconstructions of derived data for retention-driven erasures.
As none of the datasets except SmartBench contained derived data, we manually added aggregate statistics and applicable RDRs to them.
These represent real-world examples of typical derived data, e.g., the total likes of a Twitter profile, or the total sales of a supplier in TPC-H.
On the one hand, they need to be updated regularly to reflect the underlying base data.
On the other hand, it is cost-prohibitive to compute them on the fly, so minimizing the number of necessary reconstructions is desirable.

We randomly sampled 100 cells of each derived data attribute and sequentially deleted all their associated base data.
Based on the time-frame of the dataset, we varied the \textit{freq} and the grace period, $\Gamma$.
For both the $\mathbb{X}$ and Tax dataset, we set a base frequency of one day and vary the grace period between 0–23 hours.
The SmartBench dataset uses a \textit{freq} of one hour and $\Gamma$ between 0–55 minutes.
For the HotCRP and TPC-H datasets, we choose a base frequency of one week and one month, and $\Gamma$ between 0–6 and 0–30 days, respectively.
In Fig.~\ref{fig:eval:scheduling}, we depict the average number of reconstructions that we save compared to the baseline. 
We observe the effectiveness of \textsc{Scheduler} in all cases, but it differs depending on the update characteristic of the dataset.
There are three distinct patterns visible.
First, HotCRP and SmartBench are insensitive to an increase of the grace period because only base data from the same time is aggregated.
Thus, the maximal saving is reached as soon as we allow scheduling.
The actual improvement (55.8\% for HotCRP, 4.4\% for SmartBench) differ based on the number of base data cells that are aggregated into a derived cell.
This difference is also apparent between the Twitter and TPC-H dataset.
Both these datasets experience ``bursty'' updates, so initially increasing the grace period reduces the number of reconstructions significantly, but the effect flattens off.
The third update characteristic is exhibited in the Tax dataset.
It is continuously updated, so there is a steady reduction of necessary reconstructions.
Since no two updates happen at the same time, there is no benefit without a grace period.
Given the large amount of base data cells per derived data cell, the overall improvement is the largest in this dataset.

\noindent \textbf{Experiment 5.} 
In this experiment, we combine both demand-driven and retention-driven erasures.
To investigate the effect of different shares of retention-driven erasures, we employ a similar setting to Experiment~4.
We vary the fraction of profiles that are erased using \textsc{Scheduler} (based on retention-time) vs.\ demand-driven on the fly.
The demand-driven erasures are simulated by generating a random deletion time between the experiment start and the expiration time.
For the entire experiment, we allow a grace period ($\Gamma$) of one hour for Twitter and Tax, one minute for SmartBench, one day for HotCRP, and one week for TPC-H.
In both erasure methods, we adopt a time-based batching.
\begin{figure}[ht]
    \centering
    \includegraphics[width=\linewidth]{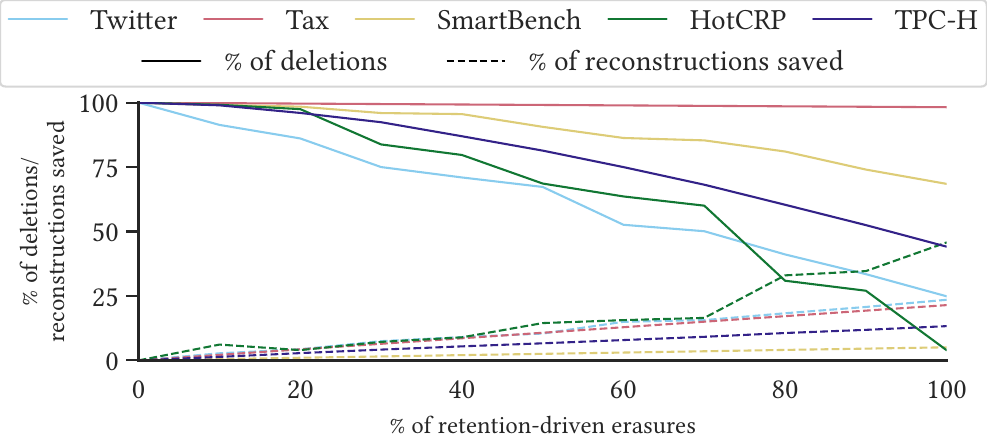}
    \Description{Overheads for fractions of demand- and retention-driven erasures shown as a graph.}
    \caption{P2E2 overheads for fractions of demand- and retention-driven erasures}
    \label{fig:eval:demand_retention}
\end{figure}

We present the average number of deleted cells and the necessary reconstructions in Fig.~\ref{fig:eval:demand_retention}.
The figure shows that 
the number of deleted cells and the number of reconstructions reduce as the share of the retention-driven erasures increases.
The improvement largely depends on the update characteristic, as explained for Experiment~4.
Thus, the HotCRP dataset profits the most, while we cannot reduce the number of deleted cells for the Tax dataset.
The variation in the amount of change depends on the number of base data cells per derived data cells.
Some derived data cells that aggregate more data are more impactful, depending on the method used to delete them.
When comparing 0\% to 100\% retention-driven erasures, we can save between 25\% (SmartBench) and 90\% (HotCRP) of deletions, and between 50\% (HotCRP) and 20\% (Twitter) of reconstructions.


\section{Related Work}
\label{sec: related_works} 
\noindent \textbf{Deletion Semantics.} Regulation-driven deletions have been studied in several contexts:  for timely deletion in  LSM-tree~\cite{Lethe},  for cascading deletes in relationship graphs~\cite{CohnDeletionSocialNetwrks}, for schema and annotations containing personal data~\cite{albab2023k9db}, for deletion in blockchains~\cite{kuperberg2020towards}, and in SQL context ~\cite{sarkar2022query} that explores extensions to specify when data should be deleted. None of these consider the role of data inference in deletion. The need for formal specification of deletion semantics for regulatory compliance has  been  discussed  in ~\cite{chakraborty2023data, BuildingDelCompDB, rupp2022leave}.

\noindent \textbf{Dependency rules.}  RDRs draw inspiration from several lines of work in databases that explored dependency frameworks. These include: specification and reasoning frameworks for functional dependencies \& denial constraints \cite{rekatsinas2017holoclean} in data cleaning \cite{rekatsinas2017holoclean} and consistent query answering \cite{bertossi2006consistent, fagin2016declarative, galhardas2001declarative, arenas1999consistent, chomicki2005minimal}, correlation constraint and causal constraints in causal databases, delta rules for  generalized reasoning for a large class of dependencies (e.g., denial constraints, correlation, and causal constraints) for deletion-based data repair ~\cite{gilad2020multiple}, and provenance/lineage dependencies using semiring structures and annotations~\cite{cheney2009provenance, Green2007, green2007update, Ujcich2018} when available. 
RDRs extend these to include aggregate dependencies (like summation and max), and cell level dependencies which are essential to define fine grained data deletion. 
While reasoning frameworks based on above dependency specifications can be exploited to identify minimal deletion sets to make a database consistent to given constraints, they do not provide mechanisms to reason with relative changes in inferences across different database states (e.g., insertion and deletion states of a data).
Discovering dependencies from data have also been explored in~\cite{pena2022fast, bleifuss2017efficient, GDPRizer, kaminsky2023discovering, bleifuss2024discovering} which RDRs can express.

\noindent \textbf{Deletion in ML.} Recent regulatory efforts, such as the AI Act~\cite{AI_Act}, raise deletion requirements in machine learning contexts. Naively deleting training data implies model retraining, which is often impractical~\cite{carlini2022membership}. Approaches such as SISA~\cite{SISA} propose partial retraining (where our retention-driven scheduling could be applied), while others apply differential privacy~\cite{DPunlearning} to bound information leakage. Extending RDRs and P2E2 to cover dependencies between data and learned models remains an open and promising direction.
\color{black}
\section{Conclusions And Future Work}
\label{sec: conclusion}
We formalize safe data erasure as Pre-insertion Post-Erasure Equivalence (P2E2), which resolves semantic ambiguity by providing strong guarantees on deleted data—filling a key gap in current systems~\cite{BuildingDelCompDB}. We implement P2E2 using Relational Dependency Rules (RDRs), developing both exact and approximate algorithms.

While P2E2 provides a principled foundation for compliant data deletion, its adoption faces two key challenges. First, data dependencies may necessitate deleting more than what is requested to be erased. However, our evaluation across five domains—including highly dependent datasets—shows that our approaches effectively reduces this overhead.  
Future extensions may incorporate selectively applying P2E2 to data subsets, weighting dependencies, domain-specific inference, or relaxed variants of P2E2 enabling flexible trade-offs between deletion cost and retention obligations 

Second, P2E2 relies on a specified set of dependencies through which deleted data can be inferred. While RDRs provide a structured way to encode known dependencies, discovering them may incur an overhead. In practice, such dependencies often exist elsewhere in the pipeline—e.g., business logic, analytics, consistency checks, or data cleaning—or can often be learned from data~\cite{GDPRizer, bleifuss2024discovering, pena2022fast}. This requirement is consistent with the “reasonableness” standard in data regulations, which calls for reasonable, cost-effective conformative measures given current technology and implementation constraints. An interesting future direction is to extend P2E2 to protect against not only specified dependencies but any potential inferences, perhaps offering weaker, probabilistic guarantees.

Beyond refining P2E2 extensions and evaluating its applicability across domains, a key direction for future work is to extend deletion guarantees to data processing pipelines, where dependencies span multiple system components.

\begin{acks}
Chakraborty was supported by a fellowship from HPI@UCI. This work was supported by NSF Grants No. 2032525, 1545071, 1527536, 1952247, 2008993, 2133391, 2245372, and 2245373.    
\end{acks} 

\bibliographystyle{ACM-Reference-Format}
\bibliography{references}

\end{document}